\pdfoutput=1
\documentclass[letterpaper,twocolumn,10pt]{article}
\usepackage{usenix-2020-09}

\usepackage{tikz}
\usepackage{amsmath}
\usepackage{graphicx}
\usepackage{soul}
\usepackage[many]{tcolorbox}
\usepackage{etoolbox}
\usepackage{dsfont}
\usepackage{mathtools,amsmath,amssymb,latexsym,amsfonts,stmaryrd}
\usepackage{pbox}
\usepackage{amsthm}
\usepackage{multirow}
\usepackage{xfrac}
\usepackage{tikz}
\usepackage{tkz-euclide}
\usepackage{hyperref}
\usepackage{mathrsfs}
\usepackage{mathpartir}
\usepackage{url}
\usepackage{syntax}
\usepackage{framed}
\usepackage{algorithm}
\usepackage[noend]{algpseudocode}
\usepackage{flushend}
\usepackage{listings}
\usepackage{moresize}
\usepackage{wrapfig}
 
\usepackage{seqsplit}
\usepackage{alltt}
\usepackage{xspace}
\usepackage{enumitem}
\usepackage{xcolor}
\usepackage{pifont}% http://ctan.org/pkg/pifont

\DeclareMathAlphabet{\mathcal}{OMS}{cmsy}{m}{n}

\usepackage{amsmath, listings, amsthm, amssymb, proof, xspace}

\usepackage{thmtools}

\declaretheoremstyle[spaceabove=\topsep,notefont=\normalfont\itshape]{mystyle}

\newcommand{\revise}[2]{{\color{red}{\ifx&#1&\else- #1\fi}} {\color{ForestGreen}{\ifx&#2&\else+ #2\fi}}}%
\renewcommand{\revise}[2]{#2}%

\usetikzlibrary{arrows,patterns, decorations.pathreplacing}

\newcommand{\Appx}{Appx.}
\newcommand{\F}{Fig.}
\newcommand{\E}{Eq.}
\newcommand{\T}{Table}
\renewcommand{\S}{Sec.}

\newcommand{\parh}[1]{\noindent\textbf{#1}}
\newcommand{\parhs}[1]{\noindent\underline{\textit{#1}}}

\newcommand{\Prop}{Proposition}
\newcommand{\Lem}{Lemma}
\newcommand{\Th}{Theorem}

\newcommand{\Df}{Definition}

\newtheorem{theorem}{Theorem}
\newtheorem{subtheorem}{Theorem}[theorem]

\newtheorem{lemma}{Lemma}
\newtheorem{definition}{Definition}
\newtheorem{proposition}{Proposition}

\newcommand{\ignore}[1]{}

\lstdefinestyle{base}{
  moredelim=**[is][\color{red}]{@}{@},
  escapeinside={<@}{@>}
}

\newcommand{\tool}{\textsc{CacheQL}\xspace}

\newcommand{\pp}{\texttt{Prime+Probe}}

\newcommand{\model}{$\mathcal{F}_{\theta}$\xspace}

\usepackage{amssymb}% http://ctan.org/pkg/amssymb
\usepackage{pifont}% http://ctan.org/pkg/pifont
\usepackage{bbding}% http://ctan.org/pkg/pifont

\newcommand\DejaVuttfamily{%
  \fontfamily{DejaVuSansMono-TLF}\selectfont }

\lstdefinestyle{base}{
  moredelim=**[is][\color{red}]{@}{@},
  escapeinside={<@}{@>}
}

\lstdefinelanguage
   [x64]{Assembler}     % add a "x64" dialect of Assembler
   [x86masm]{Assembler} % based on the "x86masm" dialect
   {morekeywords={CDQE,CQO,CMPSQ,CMPXCHG16B,JRCXZ,LODSQ,MOVSXD, %
                  POPFQ,PUSHFQ,SCASQ,STOSQ,IRETQ,RDTSCP,SWAPGS, %
                  rax,rdx,rcx,rbx,rsi,rdi,rsp,rbp, %
                  r8,r8d,r8w,r8b,r9,r9d,r9w,r9b}} % etc.

\let\oldbibliography\thebibliography
\renewcommand{\thebibliography}[1]{%
  \oldbibliography{#1}%
  \setlength{\itemsep}{0pt}%
}

\usepackage{listings}
\usepackage{fancyvrb}
\usepackage{color}
\definecolor{lightgray}{rgb}{.9,.9,.9}
\definecolor{darkgray}{rgb}{.4,.4,.4}
\definecolor{purple}{rgb}{0.65, 0.12, 0.82}
\definecolor{commentgreen}{RGB}{63,127,95}
\definecolor{pptred}{RGB}{192,0,0}
\definecolor{pptgreen}{RGB}{226,240,217}
\definecolor{pptyellow}{RGB}{255,242,204}
\definecolor{pptblue}{RGB}{222,235,247}
\definecolor{pptorg}{RGB}{244,177,131}
\definecolor{pptgreen1}{RGB}{31,78,121}
\definecolor{pptgreen2}{RGB}{0,153,153}

\newlength{\dpcircle}
\newlength{\rcircle}
\newlength{\dcircle}
\newcommand{\docircle}[4]{%
  \setlength{\dpcircle}{\dp\strutbox}%
  \setlength{\dcircle}{\dpcircle}%
  \addtolength{\dcircle}{\ht\strutbox}%
  \setlength{\rcircle}{0.5\dcircle}%
  \setlength{\unitlength}{1sp}%
  \begin{picture}(\number\dcircle,0)
    \color{#1}
    \put(\number\rcircle,\number\dpcircle){\circle*{\number\dcircle}}
    \color{#2}
    \put(\number\rcircle,\number\dpcircle){\circle{\number\dcircle}}
    \put(\number\rcircle,0){\makebox[0pt]{\textcolor{#3}{#4}}}
  \end{picture}%
}

\usepackage{threeparttable}
\colorlet{myPurple}{blue!40!red}
\definecolor{myOrange}{RGB}{255,192,0}

 \lstdefinelanguage{Solidity}{
   keywords={len,delete,int,void,payable, public, event, contract, typeof, new, true, false, catch, function, return, null, catch, switch, var, if, in, while, do, else, case, break,unsigned,int32_t,int16_t,for,define},
   keywordstyle=\color{violet}\bfseries,
   ndkeywords={State,Mem,i64,i32,i8, PC},
   ndkeywordstyle=\color{ForestGreen}\bfseries,
   identifierstyle=\color{black},
   sensitive=false,
   comment=[l]{//},
   morecomment=[s]{/*}{*/},
   commentstyle=\color{commentgreen}\ttfamily,
   stringstyle=\color{red}\ttfamily,
   morestring=[b]',
   morestring=[b]"
 }

\newcommand{\rnum}[1]{\uppercase\expandafter{\romannumeral #1\relax}}

\algnewcommand{\LeftComment}[1]{\Statex \(\triangleright\) #1}

\NewDocumentCommand{\statcirc}{ O{#2} m }{%
    \begin{tikzpicture}
    \fill[#2] (0,0) circle (0.8ex); % Fill circle with base colour (arg#2)
    \fill[#1] (0,0) -- (180:0.8ex) arc (180:0:0.8ex) -- cycle; % Fill a half circle filled with second colour (arg#1), if specified
    \end{tikzpicture}
}

\tikzset{%
    pics/sema/.style args={#1/#2/#3}{code={%
        \ifstrequal{#2}{0}{%
            \node[circle,minimum width=1.4mm,draw,fill=#1] {};
        }{%
            \tkzDefPoint(0,0){O}
            \tkzDrawSector[R,fill=#1](O,1.2mm)(90,90-#2)
            \tkzDrawSector[R,fill=#3](O,1.2mm)(90-#2,90-360)
    }
    }},
}

\newcommand{\CBrush}{\textcolor[RGB]{84,130,53}{\Checkmark}}
\newcommand{\XBrush}{\textcolor[RGB]{176,35,24}{\XSolidBrush}}
\newcommand{\TriUp}{\textcolor[RGB]{0,112,192}{$\Diamond$}}

\newcommand{\cBrush}{\textcolor[RGB]{84,130,53}{\ding{51}}}
\newcommand{\xBrush}{\textcolor[RGB]{176,35,24}{\ding{55}}}

\newcommand{\Circ}[2][black]{{\footnotesize\docircle{#1}{white}{white}{#2}}}
\newcommand{\Line}[2]{\hyperref[#2]{\textcolor{darkgray}{L\texttt{#1}}}}
\newcommand{\cirC}[2][white]{{\footnotesize\docircle{#1}{black}{black}{#2}}}
\newcommand{\Type}[1]{\cirC{\textbf{\texttt{#1}}}}
\newcommand{\Taint}[1]{\textcolor{pptred}{\texttt{#1}}}

\newcommand{\mydiamond}[1]{%
  \sbox0{$\lozenge$}%
  \usebox0\kern-.5\wd0\clap{\raisebox{.1ex}{\scalebox{.7}[1]{#1}}}\kern.5\wd0%
}

\newcommand{\Boxnum}[1]{\fbox{{\footnotesize \textit{\textbf{#1}}}}}

\let\OLDthebibliography\thebibliography
\renewcommand\thebibliography[1]{
  \OLDthebibliography{#1}
  \setlength{\parskip}{0pt}
  \setlength{\itemsep}{0pt plus 0.3ex}
}

\begin{document}

% \onecolumn
% \input{summary}
% \twocolumn

\date{}

\title{\tool: Quantifying and Localizing Cache Side-Channel Vulnerabilities in Production Software
\thanks{The extended version of the USENIX Security 2023 paper~\cite{yuan2023cacheql}.}}

\author{
{\rm Yuanyuan Yuan, Zhibo Liu, Shuai Wang\thanks{Corresponding author.}}\\
The Hong Kong University of Science and Technology\\
\textit{\{yyuanaq, zliudc, shuaiw\}@cse.ust.hk}
}

\twocolumn
\maketitle

\begin{abstract}
  
Cache side-channel attacks extract secrets by examining how victim software
accesses cache. To date, practical attacks on cryptosystems and media libraries
are demonstrated under different scenarios, inferring secret keys and
reconstructing private media data such as images.

This work first presents eight criteria for designing a full-fledged detector
for cache side-channel vulnerabilities. Then, we propose \tool, a novel detector
that meets all of these criteria. \tool\ precisely quantifies information leaks of
binary code, by characterizing the \textit{distinguishability} of logged side
channel traces. Moreover, \tool\ models leakage as a cooperative game, allowing
information leakage to be precisely distributed to program points vulnerable to
cache side channels. \tool\ is meticulously optimized to analyze whole side
channel traces logged from production software (where each trace can have
millions of records), and it alleviates randomness introduced by cryptographic
blinding, ORAM, or real-world noises.
  
Our evaluation \textit{quantifies} side-channel leaks of production cryptographic
and media software. We further \textit{localize} vulnerabilities reported by
previous detectors and also identify a few hundred \textit{new} leakage sites in
recent OpenSSL (ver. 3.0.0), MbedTLS (ver. 3.0.0), Libgcrypt (ver. 1.9.4). Many
of our localized program points are within the pre-processing modules of cryptosystems,
which are not analyzed by existing works due to scalability. We also
localize vulnerabilities in Libjpeg (ver. 2.1.2) that leak privacy about input
images.

\end{abstract}

\section{Introduction}
\label{sec:introduction}

Cache side channels enable confidential data leakage through shared data and
instruction caches. Attackers can recover program secrets like secret keys and
user inputs by monitoring how victim software accesses cache units. Exploiting
cache side channels has been shown particularly effective for cryptographic
systems such as AES, RSA, and
ElGamal~\cite{goldreich1996software,tromer2010efficient}. Recent attacks show
that private user data including images and text can be
reconstructed~\cite{xu2015controlled,hahnel2017high,yuan2022automated}.

Both attackers and software developers are in demand to quantify and localize
software information leakage. It is also vital to precisely distribute
information leaks toward each vulnerable program point, given that exploiting
program points that leak more information can enhance an attacker's success
rate. Developers should also prioritize fixing the most vulnerable program
points. Additionally, cyber defenders are interested in assessing subtle
information leaks over cryptosystems already hardened by mitigation
techniques (e.g., blinding).
Nevertheless, most existing cache side channel detectors focus exclusively on
qualitative analysis, determining whether programs are vulnerable without
\textit{quantifying} information that these flaws may
leak~\cite{wang2017cached,wang2019caches,weiser2018data,brotzman2019casym,yuan2022automated}.
Given the complexity of real-world cryptosystems and media libraries, scalable,
automated, and precise vulnerability localization is lacking. As a result, developers
may be likely reluctant (or unaware) to remedy vulnerabilities discovered by existing
detectors. As shown in our evaluation (\S~\ref{sec:evaluation}), attack vectors
in production software are underestimated. 

This work initializes a comprehensive view on detecting cache side-channel
vulnerabilities. We propose \textit{eight criteria} to design a full-fledged
detector. These criteria are carefully chosen by considering various important
aspects like scalability. Then, we propose \tool, an automated detector for
production software that meets all eight criteria. \tool\ quantifies information
leakage via mutual information (MI) between secrets and side channels. \tool\
recasts MI computation as evaluating conditional probability (CP),
characterizing \textit{distinguishability} of side channel traces induced by
different secrets. This re-formulation largely enhances computing efficiency and
ensures that \tool's quantification is more precise than existing works. It also
principally alleviates \textit{coverage issue} of conventional dynamic methods.

We also present a novel vulnerability localization method, by formulating
information leak via a side channel trace as \textit{a cooperative game} among
all records on the trace. Then, Shapley value~\cite{shapley201617}, a
well-established solution in cooperative game theory, helps to localize program
points leaking secrets. We rely on domain observations (e.g., side channel
traces are often sparse) to reduce the computing cost of Shapley value
from $\mathcal{O}(2^{N})$ to roughly constant with nearly no loss in
precision.\footnote{$N$, the length of a side channel trace, reaches 5M in
OpenSSL 3.0 RSA.}
\tool\ directly analyzes binary code, and captures both explicit and implicit
information flows. \tool\ analyzes \textit{entire} execution traces (existing
works require traces to be cut to reduce complexity) and overcomes
``non-determinism'' introduced by noises or hardening techniques (e.g.,
cryptographic blinding, ORAM~\cite{goldreich1996software}). 

We evaluate \tool\ using production cryptosystems including the latest
versions (by the time of writing) of OpenSSL, Libgcrypt and MbedTLS. We also
evaluate Libjpeg by treating user inputs (images) as privacy. To mimic
debugging~\cite{wang2017cached}, we collect memory access traces of target software using Intel
Pin as inputs of \tool.\footnote{Using Intel Pin to log memory access
traces is a common setup in this line of works. \tool, however, is not specific
to Intel Pin~\cite{pin}.} We also mimic automated real attacks in userspace-only scenarios,
where highly noisy side channel logs are obtained via
\pp~\cite{tromer2010efficient} and fed to \tool. \tool\ analyzed 10,000 traces
in 6 minutes and found hundreds of bits of secret leaks per software.
These results confirm \tool's ability to pinpoint all known vulnerabilities
reported by existing works~\cite{wang2019caches,weiser2018data}
and quantify those leakages. \tool\ also discovers hundreds of unknown
vulnerable program points in these cryptosystems, spread across hundreds of
functions never reported by prior works. Developers promptly confirmed
representative findings of \tool. Particularly, despite the adoption of
constant-time paradigms to harden sensitive components, cryptographic software is not
fully constant-time, whose non-trivial secret leaks are found and quantified by
\tool. \tool\ reveals the \textit{pre-process} modules, such as key
encoding/decoding and BIGNUM initialization, can leak many secrets and affect
\textit{all} modern cryptosystems evaluated. In summary, we have the
following contributions:

\begin{itemize}[noitemsep,topsep=0pt]
% \begin{itemize}
  \item  We propose eight criteria for systematic cache side-channel
  detectors, considering various objectives and restrictions. We design \tool,
  satisfying all of them;

  \item \tool\ reformulates mutual information (MI) with conditional probability
  (CP), which reduces the computing error and cost efficiently. It then estimates
  CP using neural network (NN). Our NN can properly handle lengthy side channel
  traces and analyze secrets of various types. Moreover, it does \textit{not}
  require manual annotations of leakage in training data;

  \item \tool\ further uses Shapley value to localize program points leaking
  secrets by simulating leakage as a cooperative game. With domain-specific
  optimizations, Shapley value, which is computational infeasible, is
  calculated with a nearly constant cost;

  \item \tool\ identifies subtle leaks (even with RSA blinding enabled), and its
  correctness has theoretical guarantee and empirical supports. \tool\ also
  localizes all vulnerable program points reported by prior works and hundreds
  of unknown flaws in the latest cryptosystems. Our representative findings
  are confirmed by developers. It illustrates the general concern that BIGNUM
  and pre-processing modules are largely leaking secrets and undermining recent
  cryptographic libraries.
\end{itemize}

\noindent \textbf{Research Artifact.}~To support follow-up research, we release the
code, data, and all our findings at
\url{https://github.com/Yuanyuan-Yuan/CacheQL}~\cite{snapshot}.

\section{Background \& Motivating Example}
\label{sec:example}
% \vspace{-4pt}

\parh{Application Scope.}~\tool\ is designed as a \textit{bug detector}. It
shares the same design goal with previous detectors~\cite{wang2017cached,
cacheaudit, wang2019caches, weiser2018data, jan2018microwalk}, whose main
audiences are developers who aim to test and ``debug'' software. \tool\ is
incapable of synthesizing proof-of-concept (PoC) exploits and is hence incapable
of launching real attacks. In general, exploiting cache side channels in the
real world is often a multi-step procedure~\cite{liu2016cache} that involves
pre-knowledge of the target systems and manual efforts. It is challenging, if
not impossible, to fully automate the process. For instance, exploitability may
depend on the specific hardware
details~\cite{liu2014random,liu2016cache,yarom2017cachebleed}, and in cloud
computing, the success of co-residency attacks denotes a key pre-condition of
launching exploitations~\cite{zhang2011homealone}. These aspects are not
considered by \tool\ which performs software analysis.
Given that said, we evaluate \tool\ by quantifying information leaks over side
channel traces logged by standard \pp\ attack, and as a proof of concept, we
extend \tool\ to reconstruct secrets/images with reasonable quality over logged
traces (see details in \Appx~\ref{appx:reconstruction}). We believe these
demonstrations show the potential and extensibility of \tool\ in practice.
%Nevertheless, we do \textit{not} take credit that \tool\ is launching ``real
%exploits.'' CacheQL is not designed for exploitations.

\smallskip
\parh{Threat Model.}~Aligned with prior works in this
field~\cite{bao2021abacus,cacheaudit,wang2017cached,wang2019caches,brotzman2019casym},
we assume attackers share the same hardware platforms with victim software.
Attackers can observe cache being accessed when victim software is running.
Attackers can log all cache lines (or other units) visited by the victim
software as a side channel trace~\cite{liu2016cache,yarom2017cachebleed}. 

Given a program $g$, we define the attacker's observation, a side channel trace,
as $o \in O$ when $g$ is executing $k \in K$. $O$ and $K$ are sets of all
observations and secrets. $K$ can be cryptographic keys or user private inputs like
photos. We consider a ``debug'' scenario where developers measure leakage when
$g$ executes $k$. Aligned with prior
works~\cite{weiser2018data,jan2018microwalk}, we assume that developers can
obtain noise-free $o$, e.g., $o$ is execution trace logged by Pin. We also
assume developers are interested in assessing leaks under real attacks. Indeed,
OpenSSL by default only accepts side channel reports exploitable in real
scenarios~\cite{opensslpatch}. We thus also launch standard \pp\ attack to log
cache set accesses. We aim to quantify information in $k$ leaked via $o$. We
also analyze leakage distribution across program points to localize flaws.
Developers can prioritize patching vulnerabilities leaking more information.

\smallskip
\parh{Two Vulnerablities: Secret-Dependent Control Branch and Data Access.}~Our
threat model focuses on two popular vulnerability patterns that are analyzed and
exploited previously, namely, secret-dependent control branch (SCB) and
secret-dependent data access (SDA)~\cite{7163050, cacheaudit,
doychev2017rigorous, wang2017cached, wang2019caches, brotzman2019casym,
bao2021abacus}. SDA implies that memory access is influenced by secrets, and
therefore, monitoring which data cache unit is visited may likely reveal
secrets~\cite{yarom2017cachebleed}. SCB implies that program branches are
decided by secrets, and monitoring which branch is taken via cache may likely
reveal secrets~\cite{7163050}. \tool\ captures both SCB and SDA, and it models
secret information flow. That is, if a variable $v$ is influenced (``tainted'')
by secrets via either explicit or implicit information flow, then control flow
or data access that depends on $v$ are also treated as SCB and SDA. The
definition of SDA/SCB is standard and shared among previous
detectors~\cite{wang2017cached,cacheaudit,wang2019caches,weiser2018data,jan2018microwalk}. 

\begin{figure}[!ht]
  \centering
  % \vspace{-8pt}
  \includegraphics[width=1.02\linewidth]{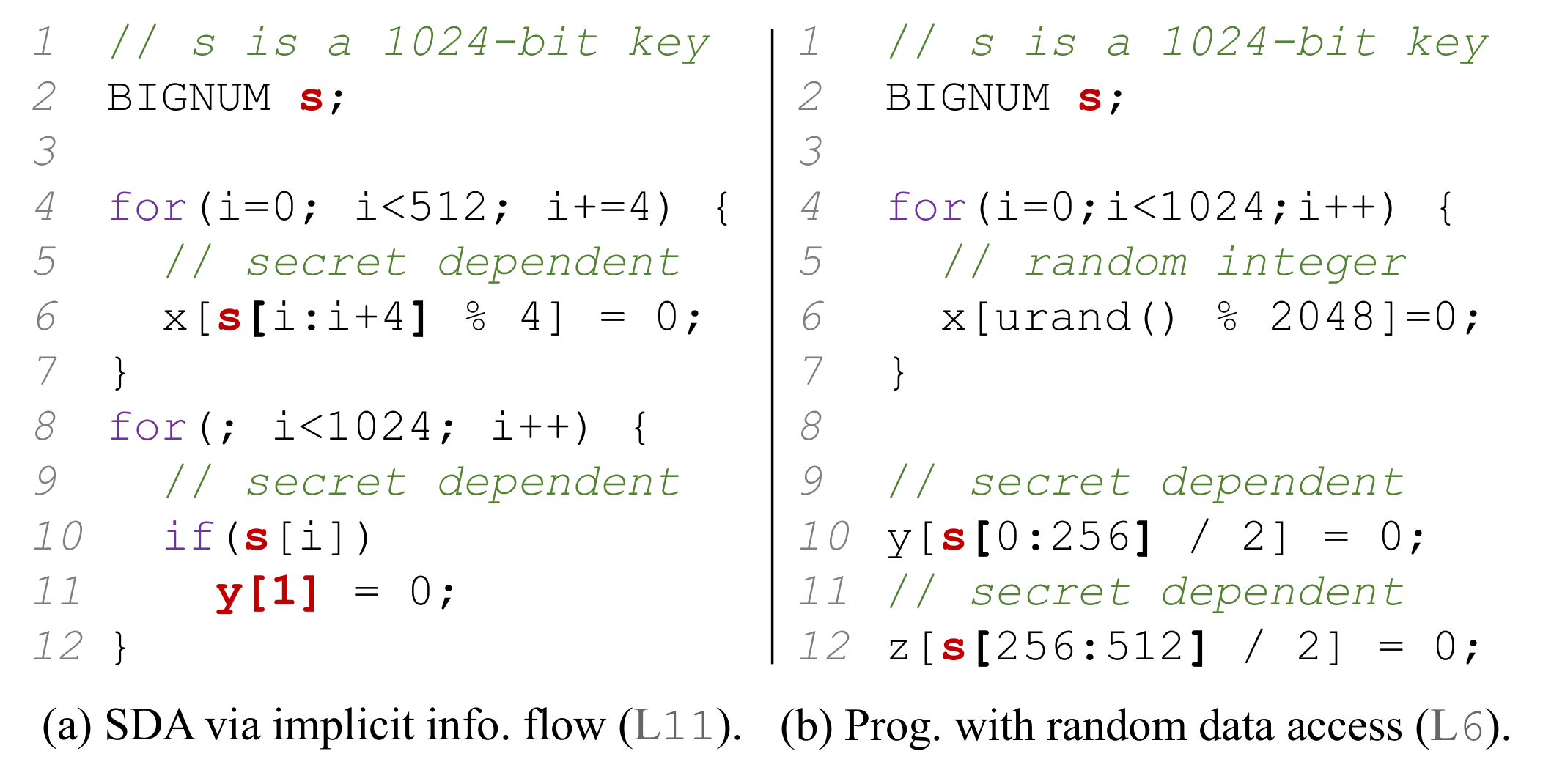}
  \vspace{-25pt}
  \caption{Two pseudocode code of secret leakage. The secrets are 1024-bit keys.
  $\texttt{s[i:j]}$ are bits between the $i$-th (included) and $j$-th bit
  (excluded).}
  \label{fig:demo-quant}
  % \vspace{-5pt}
\end{figure}

%\tool\ can capture explicit/implicit information flows.
\parh{Detecting SDA Using \tool.\footnote{SCB can be detected in the same way,
and is thus omitted here.}}~Consider two vulnerable programs depicted in
\F~\ref{fig:demo-quant}. In short, two program points in
\F~\hyperref[fig:demo-quant]{1(a)} have 128 (\Line{6}{fig:demo-quant}) and 512
(\Line{11}{fig:demo-quant}) memory accesses that are secret-dependent (i.e.,
SDA). Developer can use Pin to log one memory access trace $o$ when executing
\F~\hyperref[fig:demo-quant]{1(a)}, and by analyzing $o$, \tool\ reports a total
leakage of 768 bits. % (see\S~\ref{subsec:problem} for quantifying leaked bits).
\tool\ further apportions the SDA leaked bits as: 1) 2 bits for each of 128 memory accesses at
\Line{6}{fig:demo-quant}, and 2) 1 bit for each of 512 memory accesses at
\Line{11}{fig:demo-quant}. 
For \F~\hyperref[fig:demo-quant]{1(b)}, two array lookups at
\Line{10}{fig:demo-quant} and \Line{12}{fig:demo-quant} depend on the secret.
Given a memory access trace $o$, \tool\ quantifies the leakage as 510 bits and
apportions 255 bits for each SDA. We discuss technical details of \tool\ in
\S~\ref{sec:mi}, \S~\ref{sec:framework}, and \S~\ref{sec:local}.

\smallskip
\parh{Comparison with Existing Quantitative Analysis.\footnote{We discuss their
analysis about SDA; SCB is conceptually the
same.}}~MicroWalk~\cite{jan2018microwalk} measures information leakage via
mutual information (MI). However, we find that its output is indeed mundane
Shannon entropy rather than MI over different program execution traces, since
both key and randomness like blinding can differ traces. MicroWalk has two
computing strategies: whole-trace and per-instruction. For
\F~\hyperref[fig:demo-quant]{1(b)}, MicroWalk reports 1024 leaked bits using the
whole-trace strategy. The per-instruction strategy localizes three leakage
program points, where each point leaks 1024 (\Line{6}{fig:demo-quant}), 255
(\Line{10}{fig:demo-quant}), and 255 (\Line{12}{fig:demo-quant}) bits,
respectively. However, it is clear that those 1024 memory accesses at
\Line{6}{fig:demo-quant} are decided by \textit{non-secret} randomness. Thus,
both quantification and localization are inaccurate.
Abacus~\cite{bao2021abacus} uses trace-based symbolic execution to measure
leakage at each SDA, by estimating number of different secrets (\texttt{s}) that
induces the access of different cache units. No implicit information flow is
modelled, thereby omitting to ``taint'' the memory access at
\Line{11}{fig:demo-quant} of \F~\hyperref[fig:demo-quant]{1(a)}. Abacus
quantifies leakage of \F~\hyperref[fig:demo-quant]{1(a)} over $o$ as 256 bits,
since it only finds SDA at \Line{6}{fig:demo-quant}. 

Program points may have dependencies. For instance, one branch may have its
information leaked in its parent branch, and therefore, separately adding them
together largely over-estimates the leakage:
Abacus outputs a total leakage of 413.6 bits in AES-128, despite its 128-bit
key length.
\tool\ precisely calculates the leakage as 128.0 bits
(\S~\ref{subsubsec:aes}). Also, some static
analyses~\cite{cacheaudit,doychev2017rigorous,chattopadhyay2019quantifying} have
limited scalability due to heavyweight abstract interpretation or symbolic
execution. Real-world cryptosystems and media software are complex, with millions of
records per side channel trace. In addition, they are often unable to localize
vulnerable points.

\begin{table*}[t]
    % \vspace{-5pt}
    \caption{Benchmarking criteria for side channel detectors. \CBrush, \TriUp,
      \XBrush\ denote support, partially support, and not support.}
    \label{tab:detectors}
    \centering
  \resizebox{0.95\linewidth}{!}{
    \begin{tabular}{l|c|c|c|c|c|c|c|c|c|c|c}
      \hline
                 & CacheAudit~\cite{cacheaudit,doychev2017rigorous} & CacheD~\cite{wang2017cached} & CaSym~\cite{brotzman2019casym} & CacheS~\cite{wang2019caches} & Abacus~\cite{bao2021abacus} & CHALICE~\cite{chattopadhyay2019quantifying} & DATA~\cite{weiser2018data,weiser2020big} & MicroWalk~\cite{jan2018microwalk} & CANAL~\cite{sung2018canal} & Manifold~\cite{yuan2022automated} & \tool\ \\
      \hline
      \Circ{1} & \XBrush & \XBrush & \XBrush & \XBrush & \XBrush & \XBrush & \XBrush & \XBrush & \XBrush   & \CBrush & \CBrush \\
      \hline
      \Circ{2} & \XBrush & \XBrush & \XBrush & \XBrush & \XBrush & \XBrush & \TriUp & \XBrush & \XBrush  & \XBrush & \CBrush \\
      \hline
      \Circ{3} & \XBrush & \CBrush & \XBrush & \CBrush & \CBrush & \XBrush & \CBrush & \CBrush & \XBrush  & \CBrush & \CBrush \\
      \hline
      \Circ{4} & \TriUp  & \XBrush & \XBrush & \XBrush & \TriUp  & \TriUp & \XBrush & \TriUp & \XBrush & \XBrush & \CBrush \\
      \hline
      \Circ{5} & \XBrush & \CBrush & \CBrush & \CBrush & \CBrush & \XBrush & \CBrush & \TriUp & \CBrush  & \CBrush & \CBrush \\
      \hline
      \Circ{6} & \XBrush & \XBrush & \XBrush & \XBrush & \XBrush & \XBrush & \XBrush & \XBrush & \XBrush  & \TriUp & \CBrush \\
      \hline
      \Circ{7} & \XBrush & \XBrush & \XBrush & \XBrush & \XBrush & \XBrush & \CBrush & \CBrush & \XBrush  & \CBrush & \CBrush \\
      \hline
      \Circ{8} & \CBrush & \XBrush & \XBrush & \XBrush & \XBrush & \CBrush & \TriUp & \TriUp & \CBrush & \CBrush & \CBrush \\
      \hline
    \end{tabular}
    }
    % \vspace{-5pt}
\end{table*}

% \vspace{-5pt}
\section{Related Works \& Criteria}
\label{sec:requirement}

We propose eight criteria for a full-fledged detector. Accordingly, we review
related works in this field and assess their suitability.
\S~\ref{subsec:eval-localization} empirically compares them with \tool. Also,
many studies launch cache analysis on real-time systems and estimate worst-case
execution time
(WCET)~\cite{chu2016precise,li2003accurate,li2009timing,mitra2018time}; we omit
those studies as they are mainly for measurement, not for vulnerability
detection. 

\smallskip
\parh{Execution Trace vs.~Cache Attack Logs.}~Most existing
detectors~\cite{wang2017cached,wang2019caches,brotzman2019casym} assume access
to execution traces. In addition to recording noise-free execution traces (e.g.,
via Intel Pin), considering real cache attack logs is equally important. Cryptosystem  
developers often require evidence under real-world scenarios to issue patches.
For instance, OpenSSL by default only accepts side channel reports if they can
be successfully exploited in real-world scenarios~\cite{opensslpatch}. In sum,
we advocate that a side channel detector should~\Circ{1}~\textit{analyze both
execution traces and real-world cache attack logs}. 

\smallskip
\parh{Deterministic vs.~Non-deterministic Observations.}~Deterministic
observations imply that, for a given secret, the observed side channel is fixed.
Decryption, however, may be non-deterministic due to various masking and
blinding schemes used in cryptosystems. Furthermore, techniques like
ORAM~\cite{goldreich1996software} can generate non-deterministic memory accesses
and prevent information leakage. Thus, memory accesses or executed branches may
differ between executions using one secret. Nearly all previous
works~\cite{wang2017cached,bao2021abacus,wang2019caches,brotzman2019casym} only
consider deterministic side channels, failing to analyze the protection offered
by blinding/ORAM and may overestimate leaks (not just keys, blinding/ORAM also
change side channel observations). We suggest that a detector
should~\Circ{2}~\textit{analyze both deterministic and non-deterministic
observations.} \tool\ uses statistics to quantify information leaks from
non-deterministic observations, as explained in \S~\ref{subsec:non-det}.

\smallskip
\parh{Analyze Source Code vs.~Binary.}~A detector should typically analyze
software in executable format. This allows the analysis of legacy code and
third-party libraries. More importantly, by analyzing assembly code, low-level
details like memory allocation can be precisely considered.
Studies~\cite{doychev2017rigorous,simon2018you} reveal that compiler
optimizations could introduce side channels not visible in high-level code
representations. Thus, we argue that detectors should~\Circ{3}~\textit{be able
to analyze program executables.}

\smallskip
\parh{Quantitative vs.~Qualitative.}~Qualitative detectors decide whether
software leaks information and pinpoint leakage program
points~\cite{wang2017cached,wang2019caches,brotzman2019casym}. Quantitative
detectors further quantify leakage from each software
execution~\cite{cacheaudit, chattopadhyay2019quantifying, doychev2017rigorous},
or at each vulnerable program point~\cite{bao2021abacus, jan2018microwalk}. We
argue that a detector should~\Circ{4}~\textit{deliver both qualitative and
quantitative analysis}. Developers are reluctant to fix certain vulnerabilities,
as they may believe those defects leak negligible secrets~\cite{bao2021abacus}.
However, identifying program points that leak large amounts of data can push
developers to prioritize fixing them. To clarify, though quantitative analysis
was previously deemed costly~\cite{jancar2021they}, \tool\ features efficient
quantification.

\smallskip
\parh{Localization.}~Along with determining information leaks, localizing
vulnerable program points is critical. Precise localization helps developers
debug and patch. Therefore, a detector should~\Circ{5}~\textit{localize
vulnerable program points leaking secrets.} Most static detectors struggle to
pinpoint leakage points~\cite{cacheaudit,doychev2017rigorous}, as they measure
the number of different cache statuses to quantify leakage. Trace-based analysis,
including \tool, can identify leakage instructions on the trace that can be
mapped back to vulnerable program points~\cite{wang2017cached,bao2021abacus}.

\smallskip
\parh{Key vs.~Private Media Data.}~Most detectors analyze cryptosystems to
detect key leakage~\cite{wang2017cached,cacheaudit,bao2021abacus}. Recent side
channel attacks have targeted media data~\cite{xu2015controlled,hahnel2017high}.
Media data like images used in medical diagnosis may jeopardize user privacy
once leaked. We thus advocate detectors to~\Circ{6}~\textit{analyze leakage of
secret keys and media data}. Modeling information leakage of high-dimensional
media data is often harder, because defining ``information'' contained in media
data like images may be ambiguous. \tool\ models image holistic content (rather
than pixel values) leakage using neural networks.

\smallskip
\parh{Scalability: Whole Program/Trace~vs.~Program/Trace Cuts.}~Some prior
trace-based analyses rely on expensive techniques (e.g., symbolic execution)
that are not scalable. Given that one execution trace logged from cryptosystems 
can contain millions of instructions, existing
works~\cite{wang2017cached,bao2021abacus} require to first \textit{cut} a trace
and analyze only a few functions on the cutted trace.
Prior static analysis-based works may use abstract
interpretation~\cite{cacheaudit,wang2019caches}, a costly technique with limited
scalability. Only toy programs~\cite{cacheaudit} or a few sensitive functions
are analyzed~\cite{wang2019caches,doychev2017rigorous,brotzman2019casym}. This
explains why most existing works overlook ``non-deterministic'' factors like
blinding (criterion \Circ{2}), as blinding is applied \textit{prior to} executing
their analyzed program/trace cuts. Lacking whole-program/trace analysis limits
the study scope of prior works. \tool\ can analyze a whole trace logged by
executing production software, and as shown in \S~\ref{sec:evaluation}, \tool\
identifies unknown vulnerabilities in pre-processing modules of cryptographic libraries
that are not even covered by existing works due to scalability. In sum, we
advocate that a detector should be \Circ{7}~\textit{scalable for
whole-program/whole-trace analysis.}

\smallskip
\parh{Implicit and Explicit Information Flow.}~Explicit information flow
primarily denotes secret data flow propagation, whereas implicit information
flow models subtle propagation by using secrets as code pointers or branch
conditions~\cite{schwartz10}. Considering implicit information flow is
challenging for existing works based on static analysis due to scalability. They
thus do not \textit{fully} analyze implicit information
flow~\cite{wang2017cached,wang2019caches,brotzman2019casym,bao2021abacus}. We
argue a detector should \Circ{8}~\textit{consider both implicit and explicit
information flow} to comprehensively model potential information leaks. \tool\ delivers an
``end-to-end'' analysis and identifies changes in the trace due to either
implicit or explicit information flow propagation of secrets.

\smallskip
\parh{Comparing with Existing Detectors.}~\T~\ref{tab:detectors} compares
existing detectors and \tool\ to the criteria. Abacus and MicroWalk cannot
precisely quantify information leaks in many cases, due to either lacking
implicit information flow modeling or neglecting dependency among leakage sites
(hence repetitively counting leakage). CacheAudit only infers the upper bound of
leakage. Thus, they partially satisfy \Circ{4}.
MicroWalk quantifies per-instruction MI to localize vulnerable instructions,
whose quantified leakage per instruction, when added up, should not equal
quantification over the whole-trace MI (its another strategy) due to
program dependencies.
Also, MicroWalk cannot differ randomness (e.g., blinding) with secrets.
It thus partially satisfies \Circ{5}.

DATA~\cite{weiser2018data,weiser2020big} launches trace differentiation and
statistical hypothesis testing to decide secret-dependency of an execution
trace. Similar as \tool, DATA can also analyze non-deterministic traces.
Nevertheless, we find that DATA, by differentiating traces to detect leakage,
may manifest low precision, given it would neglect secret leakage if a cryptographic
module also uses blinding. It thus partially satisfies \Circ{2}.
More importantly, DATA does not deliver quantitative analysis. 

Recent research attempts to reconstruct media data like private user photos from
side channels~\cite{yuan2022automated,kwon2020improving,wu2020remove}. In
\T~\ref{tab:detectors}, we compare \tool\ with
Manifold~\cite{yuan2022automated}, the latest work in this field. Manifold
leverages manifold learning to infer media data. Manifold learning is
\textit{not} applicable to infer secret keys (as admitted
in~\cite{yuan2022automated}): unlike media data which contain perception
constraints retained in manifold~\cite{hinton1994autoencoders}, each key bit is
sampled independently and uniformly from 0 or 1. It thus partially fulfills
\Circ{6}. \tool\ is the first to quantify information leaks over cryptographic
keys and media data.

Implicit information flow (\Circ{8}) is not tracked by most existing static
analyzers. Analyzing implicit information flow requires considering more program
statements and data/control flow propagations, which often largely increases the
code chunk to be analyzed. This introduces extra hurdles for static
analysis-based approaches. DATA and MicroWalk also do not systematically capture
implicit information flow. DATA/MicroWalk first \textit{align} traces and then
compare aligned segments, meaning that they can overlook holistic differences
(unalignment) on traces.
\tool\ satisfies \Circ{8} as it directly observes and analyzes changes in the
side channel traces. Given any information flow, either explicit or implicit,
can differ traces, \tool\ captures them in a unified manner. Nevertheless, it is
evident that the implicit information flow cannot be captured by \tool\ unless
it is covered in the dynamic traces.

\smallskip
\parh{Clarification.}~These criteria's importance may vary depending on the
situations. Having only some of these criteria implemented is not necessarily
``bad,'' which may suggest that the tool is targeted for specific use cases.
Analyzing private image leakage (\Circ{6}) may not be as important as others,
especially for cryptosystem developers. We consider image leakage because
recent works~\cite{xu2015controlled,hahnel2017high,yuan2022automated} consider
recovering private media data.
We present eight criteria for building full-fledged side-channel detectors. The
future development of detectors can refer to this paper and prioritize/select
certain criteria, according to their domain-specific need. Also, we clarify that
in parallel to research works that detect side channel leaks, another line of
approaches (i.e., static verification) aims at deriving precise, certified
guarantees~\cite{cacheaudit,daniel2020binsec}.
% Our current discussion omits this aspect.

As clarified above, \tool\ can analyze real attack logs (\Circ{1}) and media
data (\Circ{6}). However, for the sake of presentation coherence, we discuss
them in \S~\ref{sec:discussion}. In the rest of the main paper, we explain
the design and findings of \tool\ using Pin-logged traces from cryptosystems.

% \vspace{-5pt}
\section{Quantifying Information Leakage}
\label{sec:mi}

\begin{figure}[!ht]
    \centering
    % \vspace{-10pt}
    \includegraphics[width=1.00\linewidth]{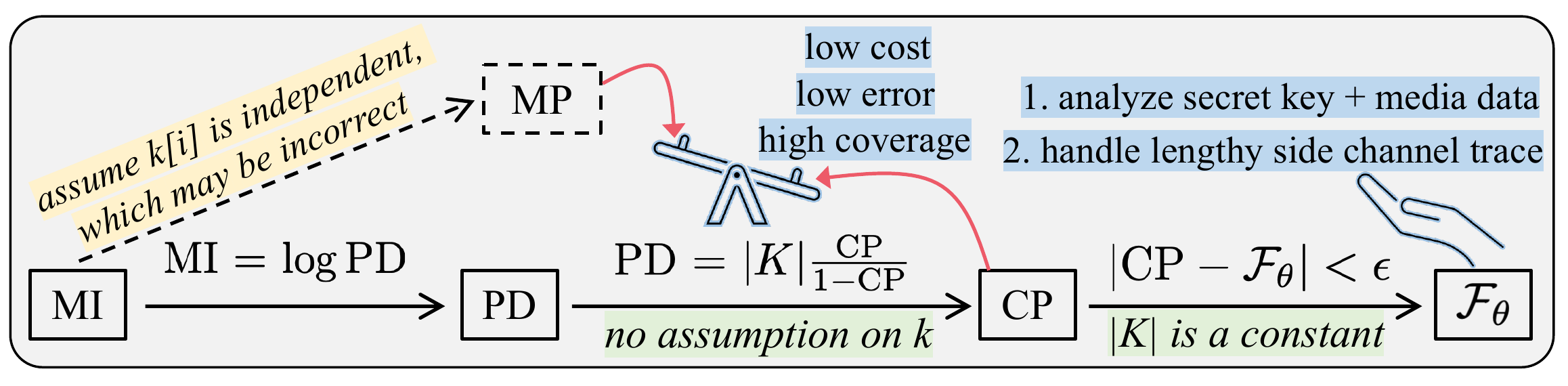}
    \vspace{-20pt}
    \caption{Overview. $|K|$ is the total number of possible keys. $|K|$ is
    assumed as known to detectors by all existing quantification tools including
    \tool.}
    \label{fig:overview}
    % \vspace{-5pt}
\end{figure}

\parh{Overview.}~This section discusses quantitative measurement of information leaks
over side channel observations. We start with preliminaries in \S~\ref{subsec:problem}.
The overview of our approach is illustrated in \F~\ref{fig:overview}.
\S~\ref{subsec:mi-pdf} introduces MI computation via Point-wise Dependence (PD).
Then, \S~\ref{subsec:pdf-est} recasts calculating PD into computing conditional
probability (CP). \tool\ employs parameterized neural networks $\mathcal{F}_{\theta}$
(see \S~\ref{sec:framework}) to estimate CP, which is carefully designed to quantify leakage of keys and private
images from extremely lengthy side channel traces. The error of estimating CP with
$\mathcal{F}_{\theta}$ is bounded by a negligible $\epsilon$. In contrast, prior
works use marginal probability (MP) to estimate MI. CP outperforms MP in terms of
lower cost, fewer errors, and better coverage, as compared in \S~\ref{subsec:pdf-est}.
In \S~\ref{subsec:non-det}, we extend the pipeline in \F~\ref{fig:overview} to handle
non-deterministic side channel traces.

\vspace{-10pt}
\subsection{Problem Setting}
\label{subsec:problem}

\begin{figure}[!ht]
    % \vspace{-5pt}
    \centering
    \includegraphics[width=1.00\linewidth]{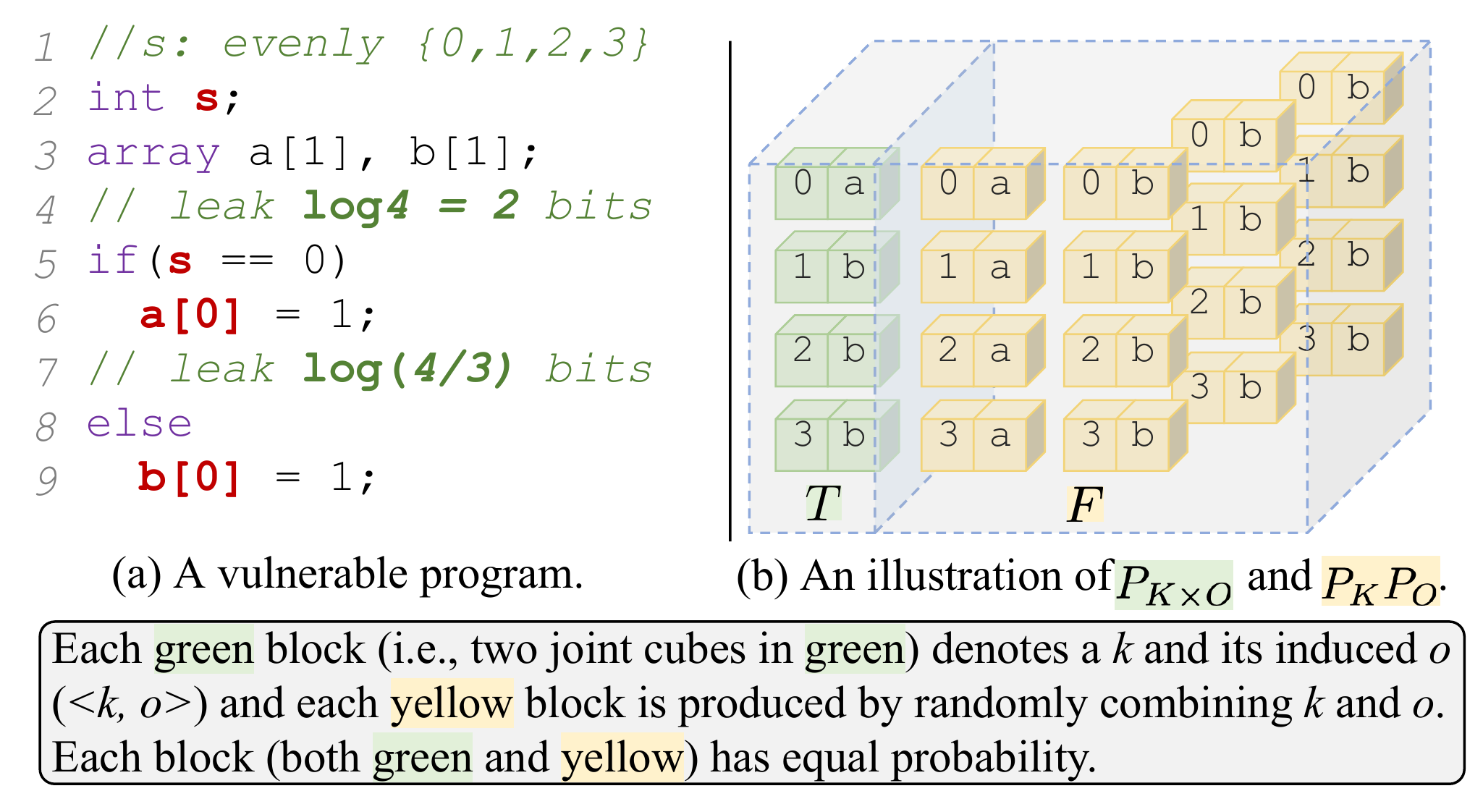}
    \vspace{-25pt}
    \caption{Quantification of side channel leaks.}
    \label{fig:demo-entropy}
    % \vspace{-10pt}
\end{figure}

In general, side channel analysis aims to infer $k$ from $o$. The information
leak of $K$ in $O$ can be defined as their MI:

% \vspace{-5pt}
\begin{equation}
    \label{equ:mi}
    {
    % \small
    I(K;O) = H(K) - H(K|O),
    }
\end{equation}
% \vspace{-10pt}

\noindent where $H(\cdot)$ denotes the entropy of an event.
According to Shannon's information theory, $I(K;O)$ describes how much
information about $K$ can be obtained by observing $O$.
% Alternatively, it measures, by giving $O$, how much uncertainty is reduced in $K$.
Consider the program in \F~\hyperref[fig:demo-entropy]{3(a)}, where the
probability of correctly guessing each $k \in K$ (i.e., $\texttt{s} \in
\{\texttt{0}, \texttt{1}, \texttt{2}, \texttt{3}\}$), without any observation,
is $\frac{1}{4}$. Thus, $H(K)$ = $-\log\frac{1}{4}$ = $2$ bits.\footnote{Given
log base 2 is used by default, the unit of information is \textit{bit}.}
Nevertheless, the observation $o$ = $a[0]$ (\Line{6}{fig:demo-entropy})
indicates that $k$ must be ``\texttt{0}'' (i.e., the probability is $1$), thus,
$H(K|o$=$a[0])$ = $-\log 1$ = $0$. Therefore, $a[0]$ leaks 2 bits of
information.
Similarly, the memory access $b[0]$ (\Line{9}{fig:demo-entropy}) leaks $\log
\frac{4}{3}$ bits of information since $H(K|o$=$b[0])$ = $- \log \frac{1}{3}$ =
$\log 3$. Ideally, a secure program should have $H(K)$ = $H(K|O)$, indicating no
information in $K$ can be obtained from $O$. We continue discussing
\F~\hyperref[fig:demo-entropy]{3(b)} in \S~\ref{subsec:pdf-est}.

\subsection{Computing MI via PD}
\label{subsec:mi-pdf}

Following \E~\ref{equ:mi}, let $k$ and $o$ be random variables whose probability
density functions (PDF) are $p(k)$ and $p(o)$. The MI $I(K;O)$ can be
represented in the following way,

% \noindent
\begin{equation}
    \label{equ:mi-int}
    {
    % \small
    \begin{aligned}
        &I(K;O) = \int\int_{K \times O} p(k,o) \log \frac{p(k, o)}{p(k)p(o)} \mathrm{d}k\mathrm{d}o \\
        &= \mathbb{E}_{P_{K \times O}} \left[ \log \frac{p(k, o)}{p(k)p(o)} \right]
        = \mathbb{E}_{P_{K \times O}} [\log c(k, o)],
    \end{aligned}
    }
\end{equation}

\noindent where $p(k,o)$ is the joint PDF of $K$ and
$O$, and $P_{K \times O}$ is the joint distribution. 
$c(k, o) = \frac{p(k, o)}{p(k)p(o)}$ denotes \textit{point-wise dependency}
(PD), measuring discrepancy between the probability of $k$ and $o$'s
co-occurrence and the product of their independent occurrences.
Accordingly, $\log c(k, o)$ denotes the point-wise mutual
information (PMI).

The MI of $K$ and $O$, by definition, is the expectation of PMI. That is,
\E~\ref{equ:mi-int} measures the dependence retained in the joint distribution
(i.e., $\langle k, o \rangle \sim P_{K \times O}$) relative to the marginal
distribution of $K$ and $O$ under the assumption of independence
(i.e., $\langle k, o \rangle \sim P_K P_O$, where $P_K$ and $P_O$ are
marginal distributions of $K$ and $O$). When $K$ and $O$ are independent, we have
$p(k,o)$ = $p(k)$$\cdot$$p(o)$ and $\frac{p(k,o)}{p(k)p(o)}$ is $1$, thus, the leakage is
$\log 1 $=$ 0$. Nevertheless, whenever $o$ leaks $k$, $k$ and $o$ should co-occur
more often than their independent occurrences, and therefore, $c(k, o) > 1$
and $\log c(k,o) > 0$.

\E~\ref{equ:mi-int} illustrates two aspects for quantitatively computing
information leakage: 1) \textbf{PMI} $\log c(k$=$k^*, o$=$o^*)$, denoting
per trace leakage for a specific $k^*$ and its corresponding $o^*$, and 2)
\textbf{MI} $I(K;O)$, denoting program-level leakage over all possible secrets
$k \in K$. To compute $I(K;O)$, we average PMI over a collection of $\langle k,o
\rangle$, where sample-mean offers an unbiased estimation for the expectation
$\mathbb{E}(\cdot)$ of a distribution~\cite{castro2009polynomial}.

\smallskip
\parh{Comparison with Prior Works.}~Abacus~\cite{bao2021abacus} launches
symbolic execution on Pin-logged execution traces. It makes a strong assumption
that $k$ is uniformly distributed, i.e., $p(k$=$k^*)$ =
$\frac{1}{|K|}$.\footnote{``Uniform distribution'' does \textit{not}
hold for image pixel values~\cite{yuan2022automated}.}  It also assumes that
each trace must be deterministic, such that $p(k$=$k^*, o$=$o^*)$ = $p(k$=$k^*)$
for a given $o^*$ and its corresponding $k^*$. This way, approximating MI in
\E~\ref{equ:mi-int} is recasted to estimating the marginal probability (MP)
$p(o$=$o^*)$. At a secret-dependent control transfer or data access point $l$,
Abacus finds all $k \in K'$ that cover $l$. The leakage at $l$ is computed as $-
\log p(o$=$o^*)$ = $-\log(\frac{|K'|}{|K|})$. Deciding $|K'|$ via constraint
solving is costly, and therefore, Abacus uses \textit{sampling} to approximate
$|K'|$. Nevertheless, estimating MP with sampling is unstable and error-prone
(see \S~\ref{subsec:pdf-est}). MicroWalk also samples $k$ to estimate MP; it
thereby has similar issues. CacheAudit quantifies program-wide leakage. Using
abstraction interpretation, it only analyzes small programs or code fragments,
and it infers only leak upper bound. \tool\ precisely computes PMI/MI via PD and
localizes flaws. We now introduce estimating PD.

\subsection{Estimating PD $c(k,o)$ via CP}
\label{subsec:pdf-est}

Because PD makes \textit{no} assumption on the secret's distribution, our
approach can infer different types of secrets (e.g., key or images). We denote
$k$ as a general representation of one secret, and for simplicity, we write $p(k
$=$ k^*)$ as $p(k^*)$ in followings. The same applies for $o$ and $o^*$. $o^*$ is
one side channel observation produced by $k^*$. However, $o^*$ may not be the only
one, given randomness like blinding can also induce different observations even
with a fixed $k^*$. We now recast computing PD over deterministic side channels
as estimating conditional probability (CP) via binary
classification~\cite{tsai2020neural}.

\subsubsection{Transforming PD to CP}
\label{subsubsec:cp}

Let $T$ depict that a pair $\langle k, o \rangle$ co-occurs (i.e.,
positive pair $\langle k, o \rangle \sim P_{K \times O}$). Let $F$ denote that
$k$ and $o$ in $\langle k, o \rangle$  occur independently (i.e., negative pair
$\langle k, o \rangle \sim P_K P_O$). Therefore, $p(k^*,o^*)$ and $p(k^*)
p(o^*)$ can be represented as the posterior PDF $p(k^*,o^* | T)$ and $p(k^*,o^*
| F)$, respectively.
According to Bayes' Theorem, PD $c(k^*,o^*)$ is re-expressed as

\vspace{-10pt}
% \noindent
\begin{equation}
    \label{equ:cond-prob}
    {
    % \small
%    c(k^*,o^*) = 
   \text{PD} = 
   \frac{p(k^*, o^*)}{p(k^*)p(o^*)} = \frac{p(k^*,o^* | T)}{p(k^*,o^* | F)}
    = \frac{p(F)}{p(T)} \frac{p(T | k^*,o^*)}{p(F | k^*,o^*)},
    }
\end{equation}

\noindent where $p(T)$ and $p(F)$ are constants (decided by the analyzed
software). Given $P_K P_O$ is produced by separating each pair in $P_{K \times
O}$ and collecting random combinations of $k$ and $o$, $\frac{p(F)}{p(T)}$
equals to $|K|$. In practice, $P_{K \times O}$ is prepared by running the
analyzed software with each $k^*$ and collecting the corresponding $o^*$. For
the program in \F~\hyperref[fig:demo-entropy]{3(a)},
\F~\hyperref[fig:demo-entropy]{3(b)} colors $P_{K \times O}$ and $P_K P_O$ in
\colorbox{pptgreen}{green} and \colorbox{pptyellow}{yellow}.
Since $\frac{p(F)}{p(T)}$ is unrelated to $k^*$ or $o^*$, $c(k^*,o^*)$ ---
representing leaked $k^*$ from $o^*$ --- is only decided by CP $p(T|k^*,o^*)$. A
larger CP indicates that more information is leaked.

\parhs{$\bullet$ Example:}~We demonstrate this transformation using
\F~\ref{fig:demo-entropy}: for $k$=``$\texttt{0}$'' and $o$=$a[0]$, fetching a
block of $\langle \texttt{0}, \texttt{a} \rangle$ from
\F~\hyperref[fig:demo-entropy]{3(b)} has a $50\%$ chance of selecting the
\colorbox{pptgreen}{green} one (in the upper-left corner). That is, CP=$p(T |
\text{``\texttt{0}''}, a[0])$=$0.5$, and therefore, $p(T |
\text{``\texttt{0}''}, a[0])$= $p(F | \text{``\texttt{0}''}, a[0])$. Since
$\frac{p(F)}{p(T)}$=$4$, \E~\ref{equ:cond-prob} yields $4 \times \frac{0.5}{0.5}
= 4$, and therefore, $\log c(k, o)$ in \E~\ref{equ:mi-int} yields $\log 4 = 2$ bits,
equaling the leakage result computed in \S~\ref{subsec:problem}.

\subsubsection{Advantages of CP vs. Marginal Probability (MP)}
\label{subsubsec:advantage}

CP captures what factors make $o^*$,
which corresponds to $k^*$, distinguishable from other $o$. By observing both
dependent and independent $\langle k, o \rangle$ pairs, \tool\ measures the
leakage via describing how the distinguishability between different $o$ is
introduced by the corresponding $k$. It is principally distinct with existing
quantitative analysis~\cite{bao2021abacus,cacheaudit,jan2018microwalk}. Abacus
and MicroWalk approximate MP $p(o^*)$ via sampling, which has the following
three limitations compared with CP. 

\smallskip
\parh{Computing Cost.}~Estimating CP is an one-time effort
over a collection of $\langle k, o \rangle$ pairs. Estimating MP, however, has to
\textit{re-perform} sampling for each $\langle k, o \rangle$. Note that the cost
for CP to estimate over the collection of $\langle k, o \rangle$ and each
re-sampling of MP is comparable. Thus, MP is much more costly.

\smallskip
\parh{Estimation Error.}~Recall that for a leakage program
point $l$, Abacus finds all $k \in K'$ that cover $l$
via constraint solving and denotes the leakage as $-\log(\frac{|K'|}{|K|})$.
Suppose it observes the first \texttt{for} loop in
\F~\hyperref[fig:demo-quant]{1(a)} has 128 consecutive accesses to $x[0]$. To
quantify the leakage, Abacus constructs the symbolic constraint
$(s[0$:$4]\%4==0) \land \ldots \land (s[508$:$512]\%4==0)$. Nevertheless,
sampling one key that satisfies this constraint has only an extremely low
probability of $(\frac{1}{4})^{128}$. That is, the MP can be presumably
underestimated when $|K'|$ is small. Thus, the leaked information can be largely
overestimated via $-\log(\frac{|K'|}{|K|})$.
MicroWalk observes $o^*$'s frequency by sampling different $k$; it thus has
similar issues. Worse, once $o^*$ is influenced by randomness like blinding, no
$o^*$ would be identical (non-replicability). Thus, it will incorrectly regard
$p(o^*)$ as $\frac{1}{|K|}$ and report $\log |K|$ leaked bits (i.e., 
equals to the key length).
In contrast, \E~\ref{equ:cond-prob} is free from this issue: even
$o^*$ is only produced by processing one or a few $k$, \tool\ directly
characterizes PD via CP $p(\cdot|k^*,o^*)$.

Overall, CP reflects: 1) the portion of records in $o^*$ affected by its
$k^*$~\cite{tsai2020neural}, and 2) to what extent $k^*$ affects each record in
$o^*$ (see \textit{\underline{Example}} below). Further, since any difference on
$o^*$, whether due to explicit or implicit information flows, contributes to
\textit{distinguishing} $o^*$ from the rest $o \in O$, \tool\ takes both
explicit and implicit flows into consideration.

\parhs{$\bullet$ Example:}~Consider the memory accesses at
\Line{10}{fig:demo-quant} and \Line{12}{fig:demo-quant} of the program in
\F~\hyperref[fig:demo-quant]{1(b)} and suppose $o^*$ is ``$y[0]$, $z[0]$''. To
estimate the leakage over $o^*$ via MP, it requires sampling keys where
$s[0$:$256]$ constitutes either 1 or 0, so do the second 256 bits, which results in a
total of 4 cases for $s[0$:$512]$. $s[0$:$512]$ has $2^{512}$ cases, denoting a
large search space. Nevertheless, CP can infer the leakage by only observing
that, the first record in $o$ increases ``1''  (i.e., distinguishable from other
$o$) whenever $s[0$:$256]$ increases 2 (with no need to simultaneously consider
$s[256$:$512]$). The same applies to the second 256 bits.

\begin{figure}[!ht]
    \centering
    % \vspace{-5pt}
    \includegraphics[width=1.03\linewidth]{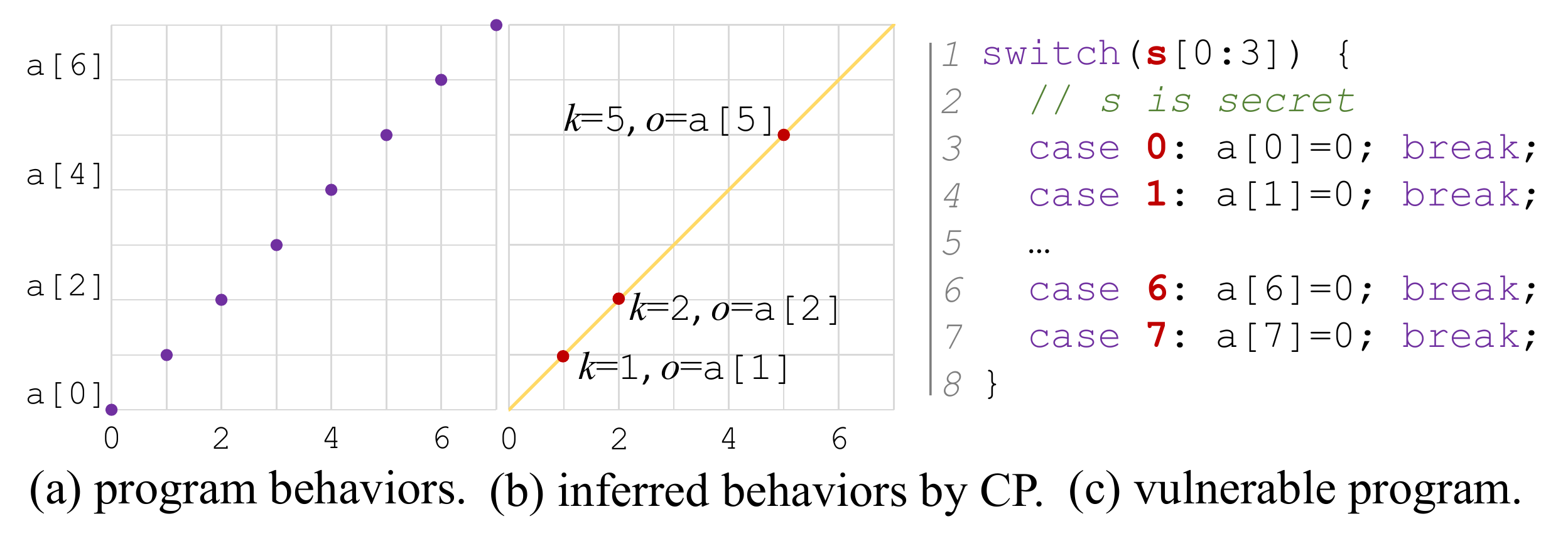}
    \vspace{-25pt}
    \caption{A schematic view of
    how the coverage issue of dynamic methods is alleviated
    via CP when \textit{quantifying} leaks.}
    \label{fig:coverage}
    % \vspace{-5pt}
\end{figure}

%\smallskip
\parh{Coverage Issue.}~\tool, by using CP, principally alleviates
the coverage issue of conventional dynamic methods.
Consider the program in \F~\hyperref[fig:coverage]{4(c)}, \tool\ can quantify
the 8 SDA \textit{without} covering all paths, since CP captures how $o$ changes
with $k$. As shown in \F~\hyperref[fig:coverage]{4(b)}, covering a few cases is
sufficient to know that $o$ increases ``one'' (e.g., $a[0] \rightarrow a[1]$)
when $k$ increases one (e.g., $0 \rightarrow 1$), thus inferring program behavior
in \F~\hyperref[fig:coverage]{4(a)}. Prior dynamic methods need to fully cover all
paths to infer the program behavior and quantify the leaks, which is hardly
achievable in practice.

\subsubsection{Obtaining CP $p(T|k, o)$ via Binary Classification}
\label{subsubsec:obtain}

We show that
performing probabilistic classification can yield CP. In particular, we employ
neural networks \model\ (parameterized by $\theta$) to classify a pair of
$\langle k, o \rangle$, whose associated confidence score is $p(T|k, o)$.
Details of \model\ are in \S~\ref{sec:framework}.

Using neural networks (NN) to estimate MI is \textit{not} our
novelty~\cite{tsai2020neural,belghazi2018mutual}. However, we deem NN as
particularly suitable for our research context for three reasons:
1) non-parametric approaches, as in~\cite{gierlichs2008mutual}, suffer from
``curse of dimensionality''~\cite{bellman1966dynamic,bengio2013representation}.
They are thus infeasible, as even the AES-128 key is 128-dimensional. NN shows
encouraging capability of handling high-dimension data (e.g., images with
thousands of dimensions). 2) Recent works show that NN can effectively process
lengthy but sparse data, including side channel traces where only a few records
out of millions are informative and leaking program
secrets~\cite{yuan2022automated,kwon2020improving,wu2020remove}.
3) It's generally vague to define ``information'' in media data. For instance, a
$64 \times 64$ image may retain the same information as a $32 \times 32$ version
from human perspective since the content is unchanged. Recent
research~\cite{yuan2022automated} shows that high-dimensional media data have
perceptual constraints which implicitly encode image ``information.'' NNs are
currently widely used to process media data and extract critical information for
comprehension.

\subsection{Handling Non-determinism}
\label{subsec:non-det}

In practice, due to hardening schemes like RSA blinding and ORAM, side channel
observations can be non-deterministic, where memory access traces may vary during
different runs despite the same key is used. As discussed in \T~\ref{tab:detectors}
(i.e.,~\Circ{2}), however, non-determinism is not properly handled in previous
(quantitative) analysis.

\smallskip
\parh{Generalizability.}~For deterministic side channels, only $k$ induces
changes of $o$. Fitting \model\ on enough $\langle k, o \rangle$ pairs from $P_{K \times O}$ and $P_K
P_O$ can capture distinguishability for quantification. In contrast, for
non-deterministic side channels, the differences between $o$ may be due to
random factors, not only $k$. Therefore, in addition to
\textit{distinguishability} between $\langle k, o \rangle$ pairs, we also need
to consider \textit{generalizability} to alleviate over-estimation caused by
random differences.

In statistics, \textit{cross-validation} is used to test generalizability. Here,
we propose a simple yet effective method by using a \textit{de-bias} term with
cross-validation to prune non-determinism in the estimated PD. We first mix
$\langle k, o \rangle$ from $P_{K \times O}$ and $P_K P_O$ and split them into
non-overlapping groups. Then, we assess if the distinguishability over one group
applies to the others.

\smallskip
\parh{PD Estimation via De-biasing.}~We first extend the PD computation in
\E~\ref{equ:cond-prob} to handle non-determinism. In \E~\ref{equ:cond-prob}, $F$
and $T$ are finite sets, and $p(F)/p(T)$ equals to $|K|$. Here, we
conservatively assume that there exist infinite non-deterministic side channels.
That is, $F$ is a set with infinite elements.
We first require $m$ positive pairs $\langle k,o \rangle \sim P_{K \times O}$,
dubbed as $T^{(m)}$. We also construct $m' \leq m^2$ negative pairs (i.e.,
$\langle k,o \rangle \sim P_K P_O$), denoted as $F^{(m')}$, by replacing $o$ (or
$k$) of a pair from $P_{K \times O}$ with that of other random pairs. This way, PD defined in
\E~\ref{equ:cond-prob} is extended in the following form:

\begin{equation}
    \label{equ:non-deter}
    {
    % \small
    \text{PD} = 
    \frac{\log |K|}{\log (m'/m)} \log \frac{p(F^{(m')})}{p(T^{(m)})}
    \frac{p(T|k^*, o^*)}{p(F|k^*, o^*)},
    }
\end{equation}

\noindent where the $p(F^{(m')})/p(T^{(m)})$ works as a \textit{de-bias} term
to assess the generalizability for non-deterministic side channels. We denote
$p(T|k^*, o^*) / p(F|k^*, o^*)$ as the \textit{leakage ratio}: a 100\% ratio
implies that all bits of the key are leaked whereas 0\% ratio implies no
leakage. Consider the following two cases:

%\smallskip
\noindent \underline{$\bullet$ $\textit{Case}_1$:}~In case the differences between samples from
$T^{(m)}$ and $F^{(m')}$ are all introduced by random noise (i.e., each $o^*$ is
independent of its $k^*$), the distinguishable factors should not be generalizable,
and the above formula yields a zero leakage. To understand this, let our neural
networks \model\ identify each pair based on random differences, which is indeed
equivalent to memorizing all pairs. This way, when it predicts the label of
$\langle k, o \rangle$, the output simply follows the frequency of $T^{(m)}$ and
$F^{(m')}$. Therefore, given an unseen pair $\langle k^*, o^* \rangle$, \model\
has $p(T|k^*, o^*) / p(F|k^*, o^*) = p(T^{(m)}) / p(F^{(m')})$, and the
estimated leakage is thus $\log
\frac{p(F^{(m')})}{p(T^{(m)})}\frac{p(T^{(m)})}{p(F^{(m')})} = \log 1 = 0$.

%\smallskip
\noindent \underline{$\bullet$ $\textit{Case}_2$:}~If $o^*$ depends on $k^*$, $p(T|k^*, o^*)$
would not merely follow the distribution of $T^{(m)}$ and $F^{(m')}$, indicating
a non-zero leakage. More importantly, de-biased by $p(F^{(m)}) / p(T^{(m')})$,
quantifying leakage using \E~\ref{equ:non-deter} \textit{only} retains differences
related to $k$. This way, we precisely quantify leakage for non-deterministic side
channels. 

\parh{Implementation Consideration.}~To alleviate randomness in each $o^*$, we collect four
observations $o_{i}^*$ by running software using $k^*$ for four times. By
classifying all $\langle k^*, o_{i}^* \rangle$ as positive pairs, \model\ is
guided to extract common characters shared by $o_{i}^*$ while neglecting
randomness in each $o_{i}^*$. Also, considering un-optimized neural networks
generally make prediction by chance (i.e., $p(T|k, o) = p(F|k, o)$), we let $m =
m'$.

% \vspace{-1pt}
\section{Framework Design}
\label{sec:framework}

\begin{figure}[!ht]
    \centering
    % \hspace{-0.5cm}
    \includegraphics[width=1.0\linewidth]{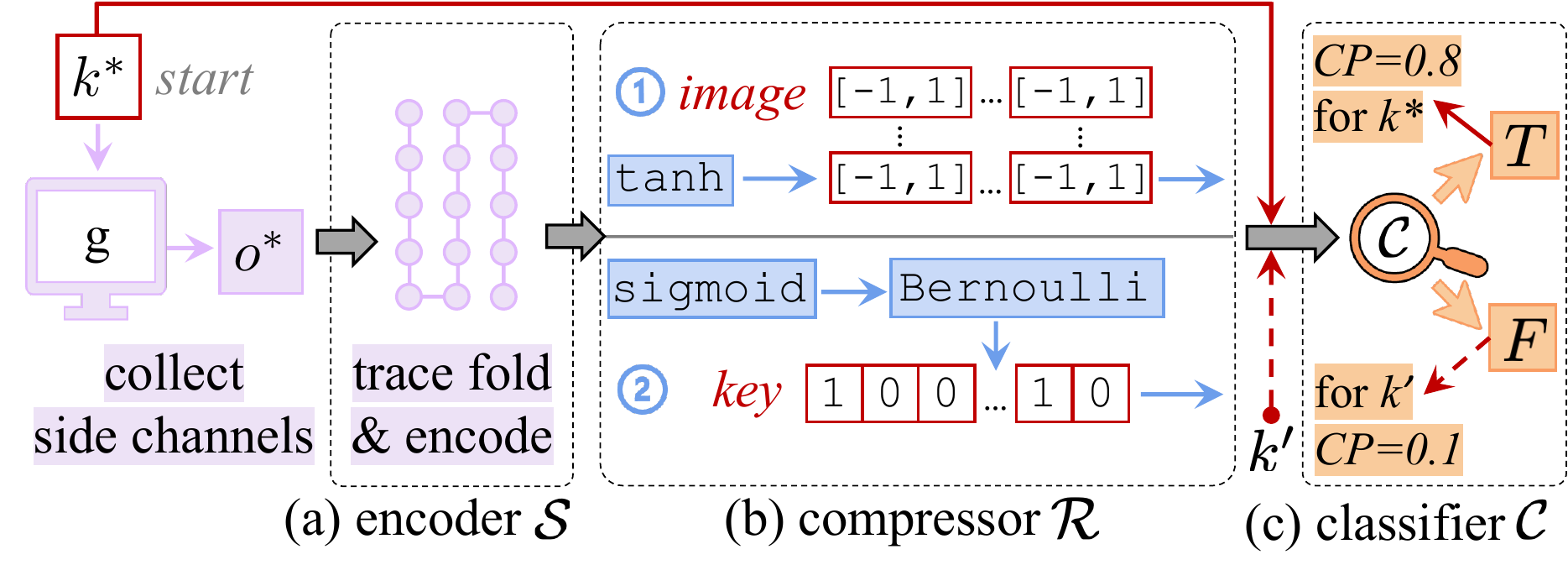}
    \vspace{-23pt}
    \caption{The framework of \tool. $\langle k^*, o^*\rangle$$\in$$P_{K \times O}$;
    it is labeled as $T$. In contrast, $\langle k', o^*\rangle$$\in$$P_K P_O$ and is
    labeled as $F$.}
    % \vspace{-2pt}
    \label{fig:framework}
\end{figure}

\F~\ref{fig:framework} shows the pipeline of \tool, including three components:
1) a sparse encoder $\mathcal{S}$ for converting side channel traces $o^*$ into
latent vectors, 2) a compressor $\mathcal{R}$ to shrink information in $o^*$,
and 3) a classifier $\mathcal{C}$ that fits the CP in \E~\ref{equ:cond-prob} via
binary classification.
We compute CP $p(T|k^*, o^*)$ using the following pipeline:

\vspace{-5pt}
\begin{equation}
  \label{equ:pipeline}
    {
    % \small
  \begin{aligned}
%    \phi(o^*)
   p(T|k^*, o^*)
   = \mathcal{F}_{\theta}(k^*, o^*) =
   \mathcal{C} (k^*, \mathcal{R}(\mathcal{S}(o^*))),
  \end{aligned}
    }
\end{equation}
\vspace{-10pt}

\noindent where parameters of these three components are jointly optimized,
i.e., $\theta = \theta_{\mathcal{S}} \cup \theta_{\mathcal{R}} \cup
\theta_{\mathcal{C}}$.

The framework takes a tuple $\langle k, o \rangle$ as input. As introduced in
\S~\ref{subsec:pdf-est}, we label a tuple $\langle k, o \rangle$ as positive if
$o$ is produced when the software is processing $k$. A tuple is otherwise
negative. In \F~\ref{fig:framework}, $\langle k^*, o^* \rangle$ is positive and
$\langle k', o^* \rangle$ is negative.

\smallskip
\parh{$\mathcal{S}$: Encoding Lengthy and Sparse Side Channel Traces.}~According
to our tentative experiments, naive neural networks perform poorly when analyzing
real-world software due to the \textit{highly lengthy} side channel traces. An
$o$, obtained via Pin or cache attacks, typically contains millions of records,
exceeding the capability of typical neural networks. ORAM can add dummy memory
accesses, often resulting in a tenfold increase of trace
length. Yuan et. al~\cite{yuan2022automated} found that side channel traces are generally
\textit{sparse}, with few secret-dependent records. It also has spatial
locality: adjacent records on a trace often come from the same or related
functions. Encoder $\mathcal{S}$ is inspired by~\cite{yuan2022automated}: to
approximate the locality, we fold the trace into a matrix (see configurations in
\S~\ref{sec:evaluation}). We employ the design in~\cite{yuan2022automated}
to construct $\mathcal{S}$ as a stack of convolutional NN layers. We find that our pipeline
effectively extracts informative features from $o$. 

\smallskip
\parh{$\mathcal{R}$: Shrinking Maximal Information.}~A side channel trace $o^*$
frequently contains information unrelated to secret $k^*$. Our preliminary study
shows that directly bridging the latent vectors (outputs of $\mathcal{S}$) to
$\mathcal{C}$ is difficult to train. This stage thereby compresses
$\mathcal{S}$'s output in an information-dense manner. We propose that
\textit{information\footnote{Only secret-related information. Non-secret
variables (e.g., public inputs) that affect $o$ are regarded as randomness
and handled as in \S~\ref{subsec:non-det}.} in $o^*$, namely $H(o^*)$, should never exceed $H(k^*)$}.
Accordingly, we apply mathematical transformations $\mathcal{R}$ to confine the
value range of $\mathcal{S}$'s outputs. We propose various transformations for
media data and secret keys; details are in \Appx~\ref{appx:r-detail}. In short,
$\mathcal{R}$ aids in retrieving secret-related information from side channels.
As demonstrated in \Appx~\ref{appx:r-detail}, \tool\ effectively extracts facial
attributes when estimating leakage of human photos.

\smallskip
\parh{$\mathcal{C}$: Optimizing Parameters via Classification.}~Let the
parameter space be $\Theta$ and $\theta \in \Theta$. To train a
neural network, we search for parameter $\theta^{\dagger} \in \Theta$ to
maximize a pre-defined objective. As shown in \E~\ref{equ:cond-prob}, we recast
leakage estimation as approximating CP $p(T|k, o)$, which is further formed
as a classification task using \model. $\theta$ is updated by gradient-guided
search in $\Theta$ to maximize the following objective: 

% \vspace{-10pt}
\begin{equation}
    \label{equ:bce}
    {
    % \small
    \begin{aligned}
        \mathbb{E}_{P_{K \times O}} [\log \mathcal{F}_{\theta}(k, o)]
        + \mathbb{E}_{P_K P_O} [\log (1 - \mathcal{F}_{\theta}(k, o))],
    \end{aligned}
    }
\end{equation}

\noindent which is a standard binary cross-entropy loss over $P_{K \times O}$
and $P_K P_O$. Overall, this loss function compares the output of \model\ to the
ground truth, and it calculates the score that penalizes \model\ based on its
output distance from the expected value. 

\parhs{$\bullet$ Example:}~Consider the program in \F~\ref{fig:demo-entropy}, in which we
have \colorbox{pptgreen}{$\langle \texttt{0}, \texttt{a} \rangle$} labeled as $T$.
To prepare $P_K P_O$, there is one
\colorbox{pptyellow}{$\langle \texttt{0}, \texttt{a} \rangle$} marked as $F$
when randomly combining $k$ and $o$ separated from pairs in $P_{K \times O}$.
Thus, $\mathcal{F}_{\theta}(\text{``\texttt{0}''}, a[0])$ is simultaneously
guided to output 1 and 0 with equal penalty. As expected, it eventually 
yields 0.5 to minimize the global penalty, which outputs a leakage of 4 bits
(since $|K|=4$) following \E~\ref{equ:optimal-trace}.

\smallskip
\parh{Computing PD.}~Let the optimized parameter be $\theta^{\dagger}$, our definition of PD in
\E~\ref{equ:cond-prob} is re-expressed in the following way to compute
point-wise information leak of $k^*$ in its derived $o^*$:

% \vspace{-5pt}
\begin{equation}
    \label{equ:optimal-trace}
    {
    % \small
    c_{\theta^{\dagger}}(k^*; o^*) = |K| \frac{\mathcal{F}_{\theta^{\dagger}}(k^*, o^*)}
    {1 - \mathcal{F}_{\theta^{\dagger}}(k^*, o^*)}
    }
\end{equation}

\noindent Furthermore, we have the following program-level information
leak assessment over $K$ and $O$.

% \vspace{-5pt}
\begin{equation}
    \label{equ:optimal-prog}
    {
    % \small
    I(K; O) = \mathbb{E}_{P_{K \times O}}
    \left[\log c_{\theta^{\dagger}} (k, o)
    \right]
    }
\end{equation}

\parh{Approximation and Correctness.}~Having access to all samples from a
distribution $P$ is difficult, if not impossible. As a common approximation, the
objective in \E~\ref{equ:bce} is instead optimized over the \textit{empirical}
distribution $P^{(n)}$ produced by $n$ samples drawn from $P$. Thus, the
estimated leakage becomes:

% \vspace{-5pt}
\begin{equation}
    {
    % \small
    \hat{I}^{(n)}_{\theta^{\dagger}}(K; O) = \mathbb{E}_{P^{(n)}_{K \times O}}
    [\log \hat{c}_{\theta^{\dagger}} (k, o)].
    }
\end{equation}

Despite we estimate MI for side channels, the skeleton for analyzing
\textit{correctness} can be adopted from prior
works~\cite{tsai2020neural,belghazi2018mutual}, since all approaches involve
optimizing parameterized neural networks. In particular, we prove that $\exists
\theta^{\dagger} \in \Theta$,

% \vspace{-10pt}
\begin{equation}
    \label{equ:appx}
    {
    % \small
    |\hat{I}^{(n)}_{\theta^{\dagger}}(K; O) - I(K; O)| \leq \mathcal{O}(
        \sqrt{\log (1 / \delta) / n}),
    }
\end{equation}

\noindent with probability at least $1 - \delta$ where $0 < \delta < 1$. We
present detailed proofs in \Appx~\ref{appx:correctness}.

% \vspace{-2pt}
\section{Apportioning Information Leakage}
\label{sec:local}

We analyze how leakage over $\langle o^*, k^* \rangle$ is apportioned among
program points. This section models information leakage as a \textit{cooperative
game} among players (i.e., program points). Accordingly, we use Shapley
value~\cite{shapley201617}, a well-developed game theory approach, to apportion
player contributions.

\parh{Overview.}~We use Shapley value (described below) to
\textit{automatically} flag certain records on a trace that contribute to leakage.
Those flagged records are \textit{automatically} mapped to assembly instructions
using Intel Pin, since Pin records the memory address of each executed instruction.
We then \textit{manually} identify corresponding vulnerable source code. We report
identified vulnerable source code to developers and have received timely confirmation
(see \Appx~\ref{appx:confirmation} for the disclosure details and their responses).
To clarify, this step is not specifically designed for Pin; users may replace Pin
with other dynamic instrumentors like Qemu~\cite{bellard2005qemu} or
Valgrind~\cite{nethercote2007valgrind}.

Shapley value decides the contribution (i.e., leaked bits) of each program point
covered on \textit{one} trace $o$. To compute the average leakage (as reported
in \S~\ref{subsec:eval-localization}), users can analyze multiple traces and
average the leaked bits at each program point.
We now formulate information leakage as a cooperative game and define leakage
apportionment as follows.

% \vspace{-3pt}
\begin{definition}[Leakage Apportionment]
    \label{def:apportion}
    Given total $n$ bits of leaked information and $m$ program points covered on
    the Pin-logged trace, an apportionment scheme allocates each program point
    $a_i$ bits such that $\sum_{i=1}^{m}a_i = n$.
\end{definition}
% \vspace{-3pt}

\parh{Shapley Value.}~We address the leakage apportionment via Shapley value.
Recall that each observation $o$ denotes a trace of logged
side channel records when target software is processing a secret $k$. Let
$\phi(o)$ be the leaked bits over one observation $o$, and let $R^o$ be the set
of indexes of records in $o$, i.e., $R^o = \{1, 2, \cdots, |o|\}$. For
\textit{all} $S \subseteq R^{o} \setminus \{i\}$, the Shapley value for the
$i$-th side channel record is formally defined as

% \vspace{-5pt}
\begin{equation}
    \label{equ:shapley}
    \pi_{i}(\phi) = \sum\limits_{S}
    \frac{|S|!(|R^o| - |S| - 1)!}{|R^o|!}[\phi(o_{S \cup \{i\}}) - \phi(o_{S})],
\end{equation}
% \vspace{-5pt}

\noindent where $\pi_{i}(\phi)$ represents the information leakage contributed
by the $i$-th record in $o$. $o_{S}$ denotes that only records whose indexes in
$S$ serve as players in this cooperative game, and accordingly $o_{R^o} = o$.
\E~\ref{equ:shapley} is based on the intuition that contribution of a player
(i.e., its Shapley value) should be decided by its marginal contribution to all
$2^{|o| - 1}$ coalitions over the remaining players. All players cooperatively
form the overall leakage $\phi(o)$.

\subsection{Computation and Optimization}
\label{subsec:optimization-sv}

The conventional procedure of deciding each player's
contribution $\pi_{i}(\phi)$ for $\phi$ requires to generate a collection of
variants $V$ over $o$, where in each variant $o_v \in V$, some players
\texttt{involved} and others \texttt{removed}~\cite{shapley201617}. In our
scenario, it is infeasible to however remove a player when estimating
leakage---removing a side channel record requires a new $\phi$. Similar
to~\cite{lundberg2017unified}, we propose to involve or remove a side channel
record from $o$ as follows:

\vspace{-5pt}
\begin{definition}[\texttt{Involved}]
    \label{def:involved}
    The $i$-th record of $o$ gets involved in $\phi(o)$ if $o[i]$ is retained.
\end{definition}

\vspace{-5pt}
\begin{definition}[\texttt{Removed}]
    \label{def:removed}
    The $i$-th record of $o$ is removed from $\phi(o)$ if $o[i]$ is reset to
    a constant, namely ``$base$''.
\end{definition}

The intuition is that, given $k$, if all records in its derived side channel
observation $o$, are set to the same constant, i.e., $o_{\emptyset} = [base,
\ldots, base]$, it's obvious that $o_{\emptyset}$ leaks no information of $k$,
namely $\phi(o_{\emptyset}) = 0$. Conversely, by gradually setting
$o_{\emptyset}[i] = o[i]$, which turns into $ o_{\{i\}}$, we finally have
$\phi(o_{R^o}) = \phi(o)$. The $base$ is $0$ in our setting for simplicity.

As stated in \S~\ref{sec:local}, computing Shapley value is costly, with
complexity $\mathcal{O}(2^{|o|})$. This is particularly challenging, since a side
channel trace $o$ frequently contains millions of records. We now propose
several simple yet highly effective optimizations which successfully reduce the
time complexity to (nearly) constant. These optimizations are based on domain
knowledge and observations about side channel traces.

\smallskip
\parh{Approximating All and Tuning Later.}~Shapley value given in
\E~\ref{equ:shapley} can be equivalently expressed as~\cite{castro2009polynomial}:

% \vspace{-10pt}
\begin{equation}
    \label{equ:shapley-sampling}
    \begin{aligned}
        \pi_{i}(\phi) &= \sum\limits_{u \in {\rm Perm}(R^o)} \frac{1}{|R^o|!}
       [\phi(o_{u^{i} \cup \{i\}}) - \phi(o_{u^{i}})] \\
       &= \mathbb{E}_{{\rm Perm}(R^o)}
       [\phi(o_{u^{i} \cup \{i\}}) - \phi(o_{u^{i}})],
    \end{aligned}
\end{equation}

\noindent where $u$ is a permutation\footnote{A set of permuted indexes. Note that the permutation
does not exchange side channel records; it provides an order
for records to get \texttt{involved}.} that assigns each position $t$ a
player indexed with $u(t)$ and ${\rm Perm}(R^o)$ is the set of all permutations
over side channel records with indexes in $R^o$. $u^{i}$ is the set of all
predecessors of the $i$-th participant in permutation $u$, e.g., if $i = u(t)$,
then $u^{i} = \{u(1), \cdots, u(t-1)\}$.

This equation transforms the computation of Shapley values into calculating the
expectation over the distribution of $u$. Each time for a randomly selected $u$,
we can calculate $\phi(o_{\{u(1), \cdots, u(j)\}})$ and $\pi_{u(j)}(\phi)$ for
all $j$ by incrementally setting $o[u(j)]$ as \texttt{involved}. Each
$\pi_{u(j)}(\phi)$ is further updated as more permutations $u$ are sampled.
From the implementation side, \E~\ref{equ:shapley} iteratively calculates
accurate Shapley values for each record (but too slow), whereas
\E~\ref{equ:shapley-sampling} approximates Shapley values for all side channel
records and tunes the values in later iterations of updates. We point out that
\E~\ref{equ:shapley-sampling} is more desirable for side channels, because
\textit{not all side channel records are correlated}. That is, updating Shapley
value for one record may not affect the results of other records (i.e.,
``Dummy Players''; see \Th~\ref{th:dummy} in \Appx~\ref{appx:sv-properties}).
% more ``Dummy Players'' (\Th~\ref{th:dummy}) will be introduced when \#coalitions increases.
Given that sample mean converges to the true expectation when \#samples
increases, $\pi_{u(j)}(\phi)$ reaches its true value when it gets convergent. As
a result, the calculation can be terminated early to reduce overhead, once the
Shapley values stay unchanged. Our empirical results show that the Shapley
values have negligible changes (i.e., the maximal difference of adjacent updates
is less than 0.5) after only tens of updates.
 
\parh{Pruning Non-Leaking Records Using Gradients.}~As discussed
in \S~\ref{sec:framework}, real-world software often generates \textit{lengthy}
and \textit{sparse} side channel records~\cite{{bao2021abacus,yuan2022automated,brotzman2019casym}}.
That is, usually only a few records in
a trace $o$ really contribute to inferring secrets, and most records are ``Null
Players'' (has no leak; see \Th~\ref{th:null} in \Appx~\ref{appx:sv-properties}) in this cooperative game. Recall that the $\phi$
is formed by neural networks, whose gradients are typically informative. Here,
we use gradients to prune Null Players before computing the standard Shapley
values. In general, neural networks characterize the influence of one input
element (i.e., one record on $o$) via gradients, and the volumes of gradients over
inputs reflect how sensitive the output is to local perturbations on these input
elements: higher volumes suggest more important elements.

We first rank all records by gradient volumes. Then, starting with the top one, we
gradually set each record as \texttt{removed}. This way, we expose Null Players,
as they are the remaining ones left when the leakage is zero. We find that, by
setting at most a few hundred records as \texttt{removed} (which is far less
than $|o|$), the leakage can be reduced to zero.

\smallskip
\parh{Batch Computations.}~The above optimizations reduce cost from
$\mathcal{O}(2^{|o|})$ to hundreds of calls to $\phi$. Further, modern hardware
offers powerful parallel computing, allowing neural networks to accept a batch
of data as inputs. Therefore, we batch the computations formed in previous steps;
eventually, with \textit{one or two batched calls} to $\phi$, whose cost is
(nearly) constant and negligible, we obtain accurate Shapley values.

\smallskip
\noindent \textbf{Error Analysis.}~Our above approximation of Shapley value is
1) \textit{unbiased}: it arrives the ground truth value with enough iterations.
It is also 2) \textit{convergent}: such that we can finish iterating whenever
the approximated value unchanged.

Let the estimated and ground truth Shapley value be $\hat{\pi}$ and $\pi$.
Previous studies have pointed out that
$\hat{\pi} \sim \mathcal{N}(\pi, \frac{\sigma^2}{m})$ where $\mathcal{N}$
is the normal distribution and $m$ is \#iterations. 
It is also proved that $\sigma^2 < \frac{(\pi_{\max} - \pi_{min})^2}{4}$ where
$\pi_{\max}$ and $\pi_{\min}$ are the maximum and minimal $\hat{\pi}$ during
all iterations~\cite{castro2009polynomial,castro2017improving}.

\smallskip
\parh{Accuracy.}~Shapley value is based on several important properties that
ensure the accuracy of localization~\cite{shapley201617}. In short, 

% \vspace{-3pt}
\smallskip
\begin{tcolorbox}[size=small]
enabled by Shapley value, leakage localization, as a cooperative game,
is precise with nearly no false negatives.
\end{tcolorbox}
% \vspace{-3pt}

The standard Shapley value ensures no false negatives (see \Th~\ref{th:null} in
\Appx~\ref{appx:sv-properties}). Nevertheless, since we trade accuracy for scalability to
handle lengthy $o$, our optimized Shapley value may have a few false negatives.
Empirically, we find that it is rare to miss a vulnerable program point, when
cross-comparing with findings of previous works~\cite{wang2019caches}.

% \vspace{-2pt}
\section{Implementation}
\label{sec:implementation}
% \vspace{-2pt}

We implement \tool\ in PyTorch (ver. 1.4.0) with about 2,000 LOC. The
$\mathcal{C}$ of \tool\ uses convolutional neural networks for images and
fully-connected layers for keys; see details in~\cite{snapshot}.
We use Adam optimizer with learning rate $0.0002$ for all models. We find that
the learning rate does not largely affect the training process (unless it is
unreasonably large or small). Batch size is 256. We ran experiments on Intel
Xeon CPU E5-2683 with 256GB RAM and a Nvidia GeForce RTX 2080 GPU. 
For experiments based on Pin-logged traces,
\S~\ref{subsubsec:scalability} presents the training time: \tool\ is generally comparable
or faster than prior tools. Experiments for \pp-logged traces
take 1--2 hours.

% \vspace{-2pt}
\section{Evaluation}
\label{sec:evaluation}
% \vspace{-2pt}

We evaluate \tool\ by answering the following research questions. \textbf{RQ1:}
What are the quantification results of \tool\ on production cryptosystems and
are they correct? \textbf{RQ2:} How does \tool\ perform on localizing side
channel vulnerabilities? What are the characteristics of these localized sites?
\textbf{RQ3:}~What are the impact of \tool's optimizations, and how does \tool\
outperform existing tools?
We first introduce evaluation setups below. 

\subsection{Evaluation Setup}
\label{subsec:setup}

\noindent \textbf{Software.}~We evaluate T-table-AES and RSA in OpenSSL 3.0.0,
MbedTLS 3.0.0, and Libgcrypt 1.9.4. We consider an end-to-end pipeline where
cryptographic libraries load the private key from files and decrypt ciphertext
encrypted from ``hello world.'' We quantify input image leaks for Libjpeg-turbo
2.1.2. We use the \textit{latest} versions (by the time of writing) of all these
software. We also assess old versions, OpenSSL 0.9.7, MbedTLS 2.15.0 and
Libgcrypt 1.6.1, for a cross-version comparison. Some of them were also analyzed
by existing works~\cite{wang2017cached,wang2019caches,bao2021abacus}. We compile
software into 64-bit x86 executable using their default compilation settings.
Supporting executables on other architectures is feasible, because \tool's core
technique is platform-independent.

\noindent \textbf{Libjpeg \& \pp.}~For the sake of presentation coherence, we
focus on cryptosystems under the in-house setting (i.e., collecting execution
traces via Pin) in the evaluation. Experiments of Libjpeg (including quantified
leaks and localized vulnerabilities) and \pp\ are in \Appx~\ref{appx:extended-eval}.

\noindent \textbf{Data Preparing \& Training.}~When collecting the data for
training/analyzing, we fix the public input and randomly sample keys
to generate side-channel traces.
For AES, we use the Linux
\texttt{urandom} utility to generate 40K 128-bit keys for estimating CP using
their corresponding side channel traces (collected via Pin or \pp). We also
generate 10K extra keys and their side channel traces to de-bias non-determinism
induced by ORAM (\S~\ref{subsubsec:quant-secure}). The same keys are used for
benchmarking AES of all cryptosystems.
For RSA, we follow the same setting but generate 1024-bit private keys using
OpenSSL.
We have no particular requirements for training data (e.g. achieving certain
coverage) --- we observe that execution flows of cryptosystems are not largely
altered by different keys, except that key values may influence certain loop
iterations (e.g., due to zero bits). We find that the execution flows of
cryptosystems are relatively more ``regulated'' than general-purpose software,
which is also noted previously~\cite{wang2017cached}. If secrets could notably
alter the execution flow, it may indicate obvious issues like timing side
channels, which should have been primarily eliminated in modern cryptosystems.

\noindent \textbf{Trace Logging.}~Pin is configured to log program memory access
traces to detect cache side channels due to SDA. We primarily consider cache
side channels via cache lines and cache banks: for an accessed memory address
\texttt{addr}, we compute its cache line and bank index as $\texttt{addr} >> 6$
and $\texttt{addr} >> 2$, respectively. We also consider SCB, where Pin logs all
control transfer destinations. Cache line/bank indexes are computed in the same
way. We clarify that cache bank conflicts are inapplicable in recent Intel CPUs;
we use this model for easier empirical comparison with prior
works~\cite{wang2017cached,bao2021abacus,wang2019caches,yuan2020private,yuan2022automated}.
Trace statistics are presented in \T~\ref{tab:trace-length}.

\parh{Ground Truth.}~To clarify, \tool\ does \textit{not} require the ground truth
of leaked bits. Rather, as discussed in \S~\ref{subsubsec:cp}, \tool\ is trained to
\textit{distinguish} traces produced when the software is processing different secrets.
The ground truth is a one-bit variable denoting whether trace $o$ is generated when
the software processing secret $k$.

\parh{Non-Determinism.}~We quantify the leaks when enabling RSA blinding
(\S~\ref{subsubsec:rsa}). We also evaluate PathOHeap~\cite{shi2020path},
a popular ORAM protocol, and consider real attack logs.

%\vspace{-5pt}
\subsection{RQ1: Quantifying Side Channel Leakage}
\label{subsec:eval-quant}

\begin{figure*}[!ht]
  \centering
  \includegraphics[width=0.93\linewidth]{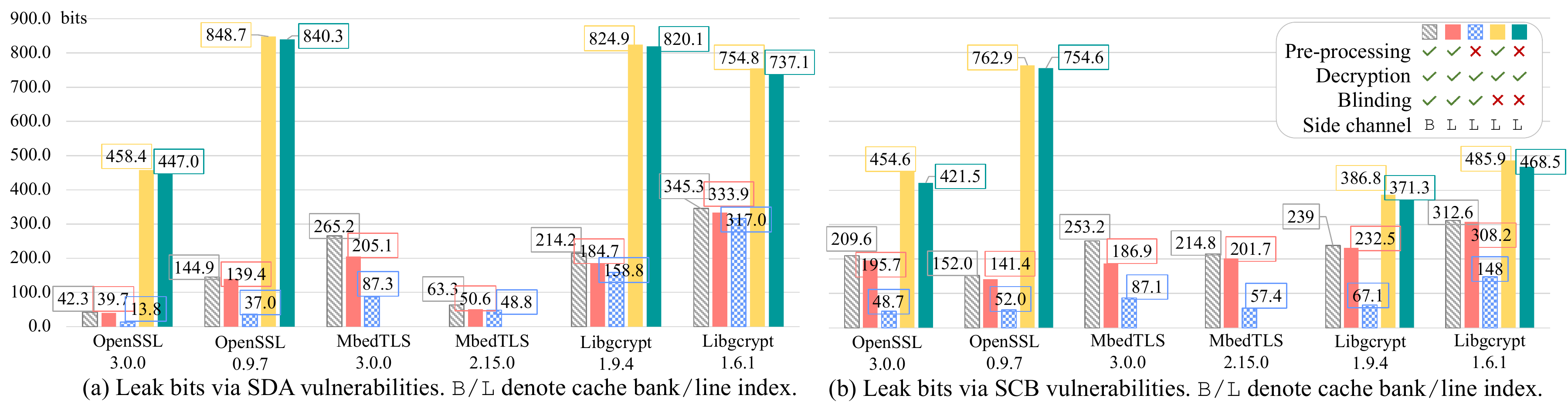}
  \vspace{-10pt}
  \caption{Leaked bits of RSA in different settings. Blinding for Libgcrypt
  1.6.1 refers to RSA optimized with CRT (see \S~\ref{subsubsec:rsa}). For cache line
  side channels (\texttt{L} in the last row of legend), we present detailed breakdown:
  enabling (\cBrush\ in the 4th row of legend) vs. disabling
  (\xBrush\ in the 4th row of legend) blinding, and with (\cBrush\ in the 2nd row) vs. w/o
  (\xBrush\ in the 2nd row) considering Pre-processing.}
  \label{fig:quant-rsa}
  \vspace{-10pt}
\end{figure*}

We report quantitative results over Pin-logged traces.
\T~\ref{tab:quant-aes} and \F~\ref{fig:quant-rsa}
summarize the quantitative leakage results computed by \tool\ regarding
different software, where a large amount of secrets are leaked across all
settings. We discuss each case in the rest of this section. Quantitative
analyses of Libjpeg and \pp\ are presented in \Appx~\ref{appx:extended-eval}.

% \vspace{-5pt}
\subsubsection{AES}
\label{subsubsec:aes}
% \vspace{-2pt}

\begin{table}[t]
  % \vspace{-5pt}
  \caption{Leaked bits of AES/AES-NI in OpenSSL/MbedTLS.}
  % \vspace{-5pt}
  \label{tab:quant-aes}
  \centering
\resizebox{0.9\linewidth}{!}{
  \begin{tabular}{c|c|c|c|c}
    \hline
       & OpenSSL 3.0.0 & OpenSSL 0.9.7 & MbedTLS 3.0.0 & MbedTLS 2.15.0 \\
    \hline
    SCB & 0             & 0             & 0             & 0              \\
    \hline
    SDA & 128.0         & 128.0         & 0             & 0              \\
    \hline
  \end{tabular}
  }
  % \vspace{-15pt}
\end{table}

The side channels of AES collected from the in-house settings are deterministic. The
SDA of AES standard T-table version can leak all key bits, but this
implementation has no SCB~\cite{bao2021abacus,wang2017cached}. These facts
serve as the ground truth for verifying \tool's quantification and localization.
MbedTLS by default uses AES-NI, which has no SDA/SCB. As shown in
\T~\ref{tab:quant-aes}, \tool\ reports no leak in it.

\tool\ reports 128 bits SDA leakage in AES-128 of OpenSSL whereas the SCB
leakage in this implementation is zero. This shows that the quantification of
\tool\ is precise. We distribute the leaked bits to program points via Shapley
value. All 128 bits are apportioned evenly toward 16 memory accesses in function
\texttt{_x86_64_AES_encrypt_compact}. Manually, we find that these 16 memory
accesses are all key-dependent table lookups.

% \vspace{-5pt}
\subsubsection{Test Secure Implementations}
\label{subsubsec:quant-secure}
% \vspace{-2pt}

\tool\ also examines secure cryptographic implementations with no leakage. For these
cases, the quantification derived from \tool\ can also be seen as their correctness
assessments. Given that said, as a dynamic method, \tool\ is for bug detection,
\textit{not} for verification.

\parh{ORAM.}~PathOHeap yields non-deterministic side channels, by randomly
inserting dummy memory accesses to produce highly lengthy traces. Since it takes
several hours to process one logged trace of RSA, we apply PathOHeap on AES from
OpenSSL 3.0.0. Overall, PathOHeap delivers provable mitigation: memory access
traces, after being processed by PathOHeap, should not depend on secrets. \tool\
reports consistent and accurate findings to empirically verify PathOHeap, as the
leaked bit is quantified as \textit{zero}.

\parh{Constant-Time Implementations.}~13 constant-time utilities~\cite{ct-utils}
from Binsec/Rel~\cite{daniel2020binsec} are evaluated using \tool. Side channel
traces from these utilities are deterministic, whose quantified leaks are also
\textit{zero}. These results empirically show the correctness of \tool's
quantification.

% \vspace{-5pt}
\subsubsection{RSA}
\label{subsubsec:rsa}
% \vspace{-2pt}

RSA blinding is enabled by default in production cryptosystems. We quantify
the information leakage of RSA with/without blinding, such that the logged
traces are \textit{non-deterministic} when blinding is on. As noted in
\S~\ref{sec:requirement} (\Circ{7}), prior works mainly focus on the decryption
fragment of RSA due to limited scalability~\cite{wang2017cached,
doychev2017rigorous, wang2019caches, bao2021abacus, brotzman2019casym}. As will
be shown, this tradeoff neglects many vulnerabilities, primarily in the
pre-processing modules of cryptographic libraries, e.g., key parsing and BIGNUM
initialization.\footnote{For simplicity, we refer to the pre-processing
functions of cryptographic libraries as Pre-processing, whereas the following
decryption functions as Decryption.} \tool\ efficiently analyzes the whole
trace, covering Pre-processing and Decryption (see \F~\ref{fig:quant-rsa}). As
an ablation, \tool\ also analyzes only Decryption, e.g., the
% \colorbox{pptgreen2}
\textcolor{pptgreen2}{green bar} in \F~\ref{fig:quant-rsa}.

\smallskip
\parh{Setup.}~Libgcrypt 1.9.4 uses blinding on both ciphertext and private keys.
We enable/disable them together. Libgcrypt 1.6.1 lacks blinding but implements
the standard RSA and another version using Chinese Remainder Theorem (CRT).
Libgcrypt 1.9.4 uses blinding in the CRT version and disables blinding in the
standard one. We evaluate these two RSA versions in Libgcrypt 1.6.1. MbedTLS
does not allow disabling blinding.

\smallskip
\parh{Results Overview.}~\F~\ref{fig:quant-rsa} shows the quantitative results.
Since cache bank only discards two least significant bits of the memory
addresses, it leaks more information than using cache line which discards six
bits. Blinding in modern cryptosystems notably reduces leakage: blinding
influences secret-dependent memory accesses, introducing non-determinism to
prevent attackers from inferring secrets.
Leakage varies across different software and variants of the same software.
Secrets are leaked via SCB and SDA to varying degrees. If blinding is disabled,
the total leak bits when considering only Decryption are close to the whole
pipeline's leakage. This is reasonable as they leak information from the same
source. With blinding enabled, leakage in Decryption is inhibited, and
Pre-processing contributes the most leakage. Though blinding minimizes leakage
in Decryption, Pre-processing remains highly vulnerable, and it is generally
overlooked previously.

\smallskip
\parh{OpenSSL.}~OpenSSL 3.0.0 has higher SCB leakage in Pre-processing with
blinding enabled. As will be discussed in \S~\ref{subsec:eval-localization},
this leakage is primarily introduced by \texttt{BN_bin2bn} and
\texttt{bn_expand2} functions, which convert key from string into BIGNUM. The
issue persists with OpenSSL 0.9.7. Moreover, compared with ver. 3.0.0,
OpenSSL 0.9.7 has more SDA (but less SCB) leakage with blinding enabled. These
gaps are also primarily caused by the \texttt{BN_bin2bn} function in Pre-processing.
We find that OpenSSL 3.0.0 skips leading zeros when converting key from string into
BIGNUM, which introduces extra SCB leakage. In contrast, OpenSSL 0.9.7 first converts
the key with leading zeros into BIGNUM and then uses \texttt{bn_fix_top} to remove
those leading zeros, causing extra SDA leakage. Also, if blinding is disabled,
OpenSSL 0.9.7 leaks approximately twice as many bits as OpenSSL 3.0.0. According to
the localization results of \tool, OpenSSL 0.9.7 has memory accesses and branch
conditions that directly depend on keys, which are vulnerable and lead to over 800
bits of leakage. We manually check OpenSSL 3.0.0 and find that most of those vulnerable
functions have been re-implemented in a constant-time way.

\smallskip
\parh{MbedTLS.}~\tool\ finds many SDA in MbedTLS 3.0.0, which primarily occurs
in the \texttt{mbedtls_mpi_read_binary} and \texttt{mbedtls_mpi_copy} functions
during Pre-processing. The problem is not severe in ver. 2.15.0. We manually
compare the two versions' Pre-processing and find that the CRT initialization
routines differ. In short, MbedTLS 3.0.0 avoids computing DP, DQ and QP (parts
of the RSA private key in CRT) and instead reads them from the PKCS1 structure,
and therefore, \texttt{mbedtls_mpi_copy} function is called for several times.
The 2.15.0 version calculates DP, DQ and QP from the private key via BIGNUM
involved functions (e.g., \texttt{mbedtls_mpi_mul_mpi}). The
\texttt{mbedtls_mpi_copy} function leaks information via SDA and SCB, whereas
the BIGNUM computation in the 2.15.0 version mainly leaks via SCB. This
difference also explains why both versions have many SCB, which are dominated by
their Pre-processing.

\smallskip
\parh{Libgcrypt.}~Libcrypt 1.9.4 has most SCB leakage in Pre-processing with
blinding enabled. Nearly all leaked bits are from the \texttt{do_vsexp_sscan}
function, which parses the key from s-expression. Decryption only leaks
negligible bits. Manual studies show that in Libgcrypt 1.9.4, most
BIGNUM-involved functions in Decryption are constant time and safe.
Nevertheless, \tool\ identifies leaks in \texttt{do_vsexp_sscan}. This
illustrates that \tool\ comprehensively analyzes production cryptosystems,
whereas developers neglect patching all sensitive functions in a constant-time
manner, enabling subtle leakages. Also, both versions have SDA leakage primarily
in the \texttt{_gcry_mpi_powm} function; this is also noted in prior
works~\cite{wang2017cached,wang2019caches,brotzman2019casym}. As aforementioned,
Libgcrypt 1.9.4 uses the standard RSA without CRT when blinding is disabled. The
1.6.1 version does not offer blinding for both the standard RSA and the CRT
version. It's obvious that the standard version leaks more than the CRT version.

\smallskip
\parh{Correctness.}~It is challenging to obtain ground truth in our evaluation.
Aside from the AES cases and secure implementations in
\S~\ref{subsubsec:quant-secure} who have the ``ground truth'' (either leaking
128 bits or zero bits) to compare with, there are several cases in RSA whose
leaked bits can be calculated manually, facilitating to assess the correctness
of \tool's quantification.

\noindent \textbf{$\text{Case}_1$:}~\texttt{BN_num_bits_word} function in
OpenSSL 0.9.7, which is first identified by CacheD~\cite{wang2017cached} and
currently fixed in OpenSSL 3.0.0, has 256 different entries depending on
secrets. It leaks $-\log \frac{1}{256}$ = $8.0$ bits, in case entries are
accessed evenly (which should be true since key bits are generated independently
and uniformly). \tool\ reports the leakage as $7.4$ bits, denoting a close
quantification.

\noindent \textbf{$\text{Case}_2$:}~\texttt{do_vsexp_sscan} function (see
\F~\ref{fig:libgcrypt}) in both versions of Libgcrypt has control branches
depending on whether a secret is greater than 10. The SCB at
\Line{2}{fig:libgcrypt} of \F~\ref{fig:libgcrypt}, in theory, leaks $-\log
\frac{1}{16} + \log \frac{1}{10}$ = $0.68$ bits of information, as the possible key
values are reduced from 16 to 10 when \Line{2}{fig:libgcrypt} is executed.
Similarly, the SCB at \Line{4}{fig:libgcrypt} leaks $-\log \frac{1}{16} + \log
\frac{1}{6}$ = $1.42$ bits. When \tool\ analyzes one trace, it apportions around
$1$ bit to each of the two records corresponding to the SCB at
\Line{2}{fig:libgcrypt} and \Line{4}{fig:libgcrypt}. We interpret that \tool\
provides accurate quantification and apportionment for this case.

% \vspace{-5pt}
\begin{tcolorbox}[size=small]
  \textbf{Answer to RQ1}: By quantifying leakage with \tool, we find that
  information leaks are prevalent in cryptosystems, even when hardening
  methods (e.g., blinding) are enabled. Most leaks reside in the pre-processing
  stage neglected by existing research. For some cases, the development of
  cryptosystems may increase the amount of leakage. The correctness of \tool\
  is empirically validated using a total of 24 instances of known bit leakages.
\end{tcolorbox}
% \vspace{-15pt}
\subsection{RQ2: Localizing Leakage Sites}
\label{subsec:eval-localization}

This section reports the leakage program points localized in RSA by \tool\ using
Shapley value. We report representative functions in \T~\ref{tab:func-cat}.
See~\cite{snapshot} for detailed reports.

When blinding is enabled, \tool localizes all previously-found
leak sites and hundreds of new ones. 

\begin{table}[t]
  % \vspace{-5pt}
  \caption{Representative vulnerable func. and their categories.}
  \label{tab:func-cat}
  \centering
  \resizebox{1.00\linewidth}{!}{
  \begin{tabular}{c|c|c|c|c|c}

    \hline
    \textbf{OpenSSL 3.0.0 }& Type & \textbf{MbedTLS 3.0.0} & Type & \textbf{Libgcrypt 1.9.4} & Type\\
    \hline
    \multirow{2}{*}{\texttt{bn_expand2}} & \Type{A}, \Type{B}, & \multirow{2}{*}{\shortstack{\texttt{mbedtls_mpi}\\\texttt{_copy}}} & \Type{A}, \Type{B}, & \multirow{2}{*}{\texttt{mul_n_basecase}} & \Type{B}, \Type{C},\\
    & \Type{C} & & \Type{C} & & \Type{D} \\
    \hline
    \multirow{2}{*}{\texttt{BN_bin2bn}} & \multirow{2}{*}{\Type{A}, \Type{C}} & \multirow{2}{*}{\shortstack{\texttt{mbedtls_mpi}\\\texttt{_read_binary}}} & \multirow{2}{*}{\Type{A}, \Type{C}} & \multirow{2}{*}{\texttt{do_vsexp_sscan}} & \multirow{2}{*}{\Type{A}, \Type{D}}\\
    & & & & & \\
    \hline
    \multirow{2}{*}{\shortstack{\texttt{BN_mod_exp}\\\texttt{_mont}}} & \Type{B}, \Type{C} & \multirow{2}{*}{\texttt{mpi_montmul}} & \Type{B}, \Type{C}, & \multirow{2}{*}{\texttt{_gcry_mpih_mul}} & \Type{B}, \Type{C},\\
    & \Type{D}, \Type{E} & & \Type{D} & & \Type{D}, \Type{E} \\
    \hline
  \end{tabular}
  }
  % \vspace{-10pt}
\end{table}

\smallskip
\parh{Clarification.}~Some leak sites localized by \tool\ are dependent, e.g.,
several memory accesses within a loop where only the loop condition depends on
secrets. To clarify, \tool\ does not distinguish dependent/independent leak
sites, because from the game theory perspective, those dependent leak sites
(i.e., players) collaboratively contribute to the leakage (i.e., the game).
Also, reporting all dependent/independent leak sites may be likely more
desirable, as it paints the complete picture of the software attack surface.
Overall, identifying \textit{independent} leak sites is challenging, and to our
best knowledge, prior works also do not consider this.
This would be an interesting future work to explore. On the other hand, vulnerabilities
identified by \tool\ are from \textit{hundreds of functions} that are not
reported by prior works, showing that the localized vulnerabilities spread
across the entire codebase, whose fixing may take considerable effort.

\subsubsection{Categorization of Vulnerabilities}
\label{subsubsec:categorization}

We list all identified vulnerabilities
in~\cite{snapshot}. Nevertheless, given the large number of (newly) identified
vulnerabilities, it is obviously infeasible to analyze each case in this paper.
To ease the comparison with existing tools that feature localization, we
categorize leak sites from different aspects. We first categorize the leak sites
according to their locations in the codebase (\Type{A} and \Type{B}). We then
use \Type{D} and \Type{E} to describe how secrets are propagated. Moreover,
since leaking-leading-zeros is less considered by previous work, we specifically
present such cases in \Type{C}.

\smallskip
\noindent \Type{A}~\ul{Leaking secrets in Pre-processing:}~Leak sites belonging
to \Type{A} occur when program parses the key and initializes relevant data
structures like BIGNUM. Note that this stage is rarely assessed by previous
static (trace-based) tools due to limited scalability; empirical results
are given in \T~\ref{tab:scalability}.

\smallskip
\noindent \Type{B}~\ul{Leaking secrets in Decryption:}~While \Type{B} is
primarily analyzed by prior static tools, in practice, they have to trade precision
for speed, omitting analysis of full implicit information flow (\Circ{8} in
\T~\ref{tab:detectors}). Therefore, their findings related to \Type{B} compose
only a small subset of \tool's findings. Also, prior dynamic tools, including
DATA and MicroWalk, are less capable of detecting \Type{B}. This is because
blinding is applied at Decryption (\Circ{2} in \T~\ref{tab:detectors}). DATA
likely neglects leak sites when blinding is enabled since it merely
differentiates logged side channel traces with key fixed/varied. MicroWalk
incorrectly regards data accesses/control branches influenced by blinding as
vulnerable. Blinding can introduce a great number of records (see
\T~\ref{tab:trace-length} for increased trace length), and MicroWalk
fails to correctly analyze all these cases.

\smallskip
\noindent \Type{C}~\ul{Leaking leading zeros:}~Besides \tool, findings
belonging to \Type{C} were only partially reported by DATA. Particularly, given
DATA is less applicable when facing blinding (noted in
\S~\ref{sec:requirement}), it finds \Type{C} only in Pre-processing, where
blinding is not enabled yet. Since \tool\ can precisely quantify (\Circ{4} in
\T~\ref{tab:detectors}) and apportion (\Circ{5}) leaked bits, it is capable of
identifying \Type{C} in Decryption; the same reason also holds for \Type{B}. At
this step, we manually inspected prior static tools and found they only
``taint'' the content of a BIGNUM, which is an array, if BIGNUM stores secrets.
The number of leading zeros, which has enabled exploitations (CVE-2018-0734 and
CVE-2018-0735~\cite{weiser2020big}) and is typically stored in a separate
variable (e.g., \texttt{top} in OpenSSL), is neglected.

\smallskip
\noindent \Type{D}~\ul{Leaking secrets via explicit information flow:}~Most
findings belonging to \Type{D} have been reported by existing static tools.
\tool\ re-discovers \textbf{all} of them despite it's dynamic.
We attribute the success to \tool's precise quantification, which recasts
MI as CP (\S~\ref{subsec:pdf-est}), and localization, where leaks are
re-formulated as a cooperative game (\S~\ref{sec:local}).

\smallskip
\noindent \Type{E}~\ul{Leaking secrets via implicit information flow:}~As
discussed above, prior static tools are incapable of fully detecting \Type{E}.
Also, many findings of \tool\ in \Type{E} overlap with that in \Type{B}. Since
DATA cannot handle blinding well (blinding is extensively used in Decryption), only
a small portion of \Type{E} were correctly identified by DATA. DATA also has the
same issue to neglect \tool's findings in \Type{D}.

In sum, static-/trace-based tools (CacheD, CacheS, Abacus) can detect \Type{B}
$\cap$ \Type{D} but cannot identify \Type{A} $\cup$ \Type{C} $\cup$ \Type{E}. As
noted in \S~\ref{sec:requirement}, MicroWalk cannot properly differ randomness
induced by blinding vs. keys, and is inaccurate for the RSA case with blinding
enabled. DATA pinpoints \Type{A} (accordingly include \Type{A} $\cap$ \Type{C})
and is less applicable for \Type{B}. \tool, due to its precise quantification,
localization, and scalability, can identify \Type{A} $\cup$ \Type{B} $\cup$
\Type{C} $\cup$ \Type{D} $\cup$ \Type{E}.

\subsubsection{Characteristics of Leakage Sites}
\label{subsubsec:characteristics}

The leakage sites exist in all stages of cryptosystems.
Below, we use case studies and the distribution of leaked bits to illustrate
their characteristics.
In short, the leaks start when parsing keys from files and initializing
secret-related BIGNUM, and persist during the whole life cycle of RSA execution.

\begin{figure}[!ht]
  %\hspace{-0.45cm}
  \vspace{-5pt}
  \centering
  \includegraphics[width=1.03\linewidth]{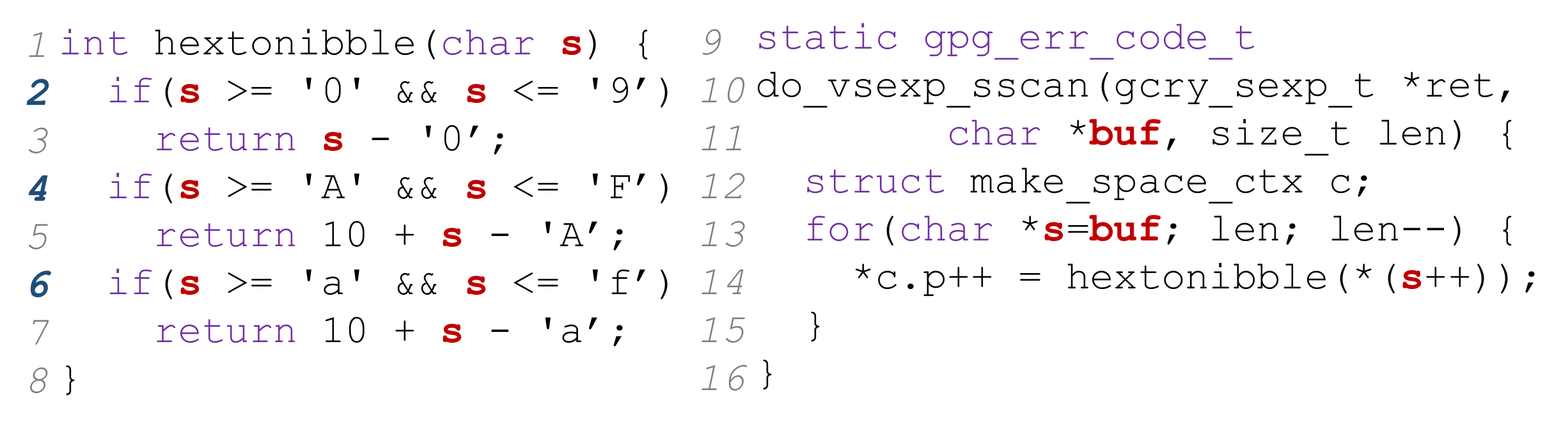}
  \vspace{-20pt}
  \caption{Simplified vulnerable program points localized in Libgcrypt
  1.9.4. This function has SCB directly depending on bits of the key.}
  \vspace{-5pt}
  \label{fig:libgcrypt}
\end{figure}

\begin{figure*}[!ht]
  \centering
  \includegraphics[width=1.0\linewidth]{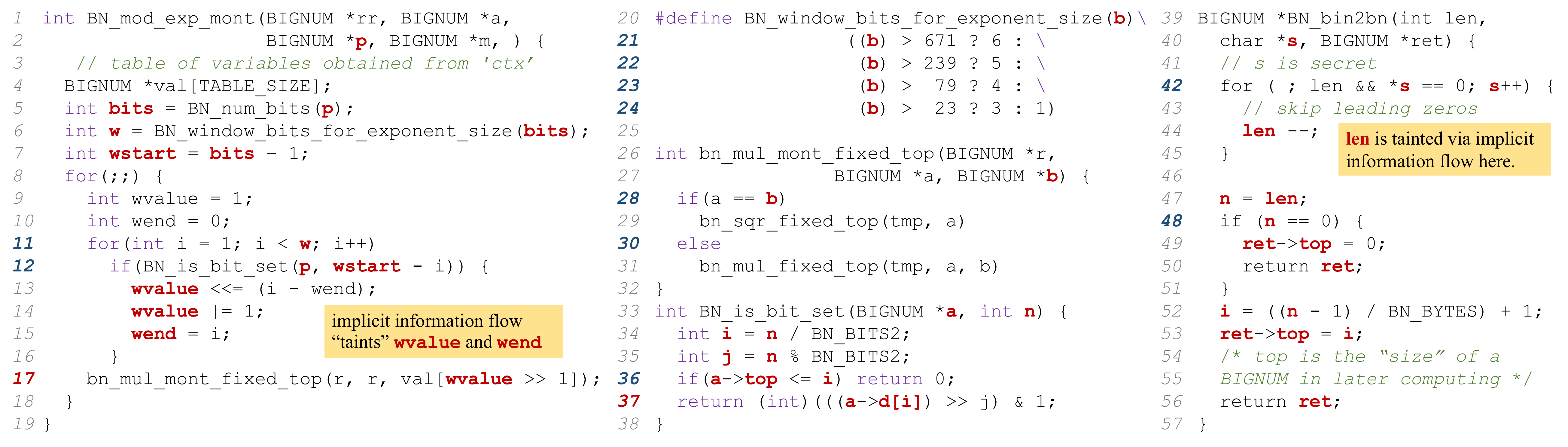}
  \vspace{-20pt}
  \caption{Vulnerable program points localized in OpenSSL 3.0.0. We mark the
  line numbers of \textcolor{pptred}{SDA vulnerabilities} and
  \textcolor{pptgreen1}{SCB vulnerabilities} found by \tool. Secrets are
  propagated from \Taint{p} to other
  \textcolor{pptred}{variables} via explicit or implicit information flow, which
  confirm each SDA/SCB vulnerability found by \tool.}
  \label{fig:openssl}
  \vspace{-5pt}
\end{figure*}

\parh{Case Study$_1$:}~\F~\ref{fig:libgcrypt} presents a case newly disclosed
by \tool, which is the key parsing implemented in Libgcrypt 1.6.1 and 1.9.4. As
discussed in \S~\ref{subsubsec:rsa} (see $\text{Case}_2$), this function has SCB
explicitly depending on the key read from files. It therefore contains
\Type{A} and \Type{D}. Similar leaks exist in other software. For instance, as
localized by \tool\ and DATA, the \texttt{EVP_DecodeUpdate} function in two
versions of OpenSSL have SDA via the lookup table \texttt{data_ascii2bin} when
decoding keys read from files.

\smallskip
\parh{Case Study$_2$:}~\F~\ref{fig:openssl} depicts the life-cycle of
BIGNUM in OpenSSL 3.0.0, including initialization and computations. We show how
secrets are leaked along the usage of BIGNUM.

\smallskip
\noindent \Boxnum{1} \texttt{BN_bin2bn}@\Line{39}{fig:openssl}: A BIGNUM is
initialized using \Taint{s} at \Line{40}{fig:openssl}, which is parsed
from the key file in the \texttt{.pem} format. A \texttt{for} loop at
\Line{42}{fig:openssl} \textit{skips leading zeros}, propagating \Taint{s} to
\Taint{len} via implicit information flow. Then, \Taint{len} is propagated to
\Taint{top} (\Line{49}{fig:openssl} or \Line{53}{fig:openssl}). Thus, future
usage of \Taint{top} clearly leaks secret.

\smallskip
\noindent \Boxnum{2} \texttt{BN_mod_exp_mont}@\Line{1}{fig:openssl}:
\texttt{BN_num_bits} is called to calculate \#bits (after excluding leading
zeros) of BIGNUM \Taint{p}. \texttt{BN_num_bits} further calls
\texttt{BN_num_bits_word} which we have discussed in \S~\ref{subsubsec:rsa}. \#bits
is stored in \Taint{bits} at \Line{5}{fig:openssl}. Later, \Taint{bits} is
propagated to \Taint{wstart} at \Line{7}{fig:openssl}.

\smallskip
\noindent \Boxnum{3} \texttt{BN_window_bits_for_exponent_size}@\Line{20}{fig:openssl}:
\Taint{w} is propagated from \Taint{bits} at \Line{6}{fig:openssl},
given control branches from \Line{21}{fig:openssl} to \Line{24}{fig:openssl}
directly depend on \Taint{b}.

\smallskip
\noindent \Boxnum{4} \texttt{BN_is_bit_set}@\Line{33}{fig:openssl}:
\Taint{top} of BIGNUM \Taint{p} directly decides the return value at
\Line{36}{fig:openssl}. Its content, namely array \Taint{d}, also sets the
return value at \Line{37}{fig:openssl}. Given \Taint{wvalue} and \Taint{wend} at
\Line{13}{fig:openssl} and \Line{15}{fig:openssl} are updated according to the
return value of \texttt{BN_is_bit_set}, they are thus implicitly propagated.

\smallskip
\noindent \Boxnum{5} \texttt{bn_mul_mont_fixed_top}@\Line{26}{fig:openssl}:
The access to array \texttt{val} at \Line{17}{fig:openssl} is indexed with
\Taint{wvalue}, and therefore, it induces SDA. Variable \Taint{b} at
\Line{27}{fig:openssl} is also propagated via \Taint{wvalue}, and the \texttt{if}
branch at \Line{28}{fig:openssl} thus introduces SCB.

Overall, \Boxnum{1} executes at Pre-processing and is only detected
by DATA and \tool. It has both explicit (\Line{47}{fig:openssl})
and implicit (\Line{42}{fig:openssl}) information flow. Thus, it has
\Type{A}~\Type{C}~\Type{D}~\Type{E}. Similarly, \Boxnum{2} contains
\Type{B}~\Type{C}~\Type{D}~\Type{E}. Both \Boxnum{3} and \Boxnum{4}
have \Type{B}~\Type{C}~\Type{D}. \Boxnum{5} only has \Type{B}~\Type{D}.
Among the leak sites discussed above, only five SCB at
\Line{21}{fig:openssl}-\Line{24}{fig:openssl} and
\Line{36}{fig:openssl} are detected by previous static tools; remaining ones
are newly reported by \tool.

\begin{figure*}[!ht]
  \centering
  \includegraphics[width=0.65\linewidth]{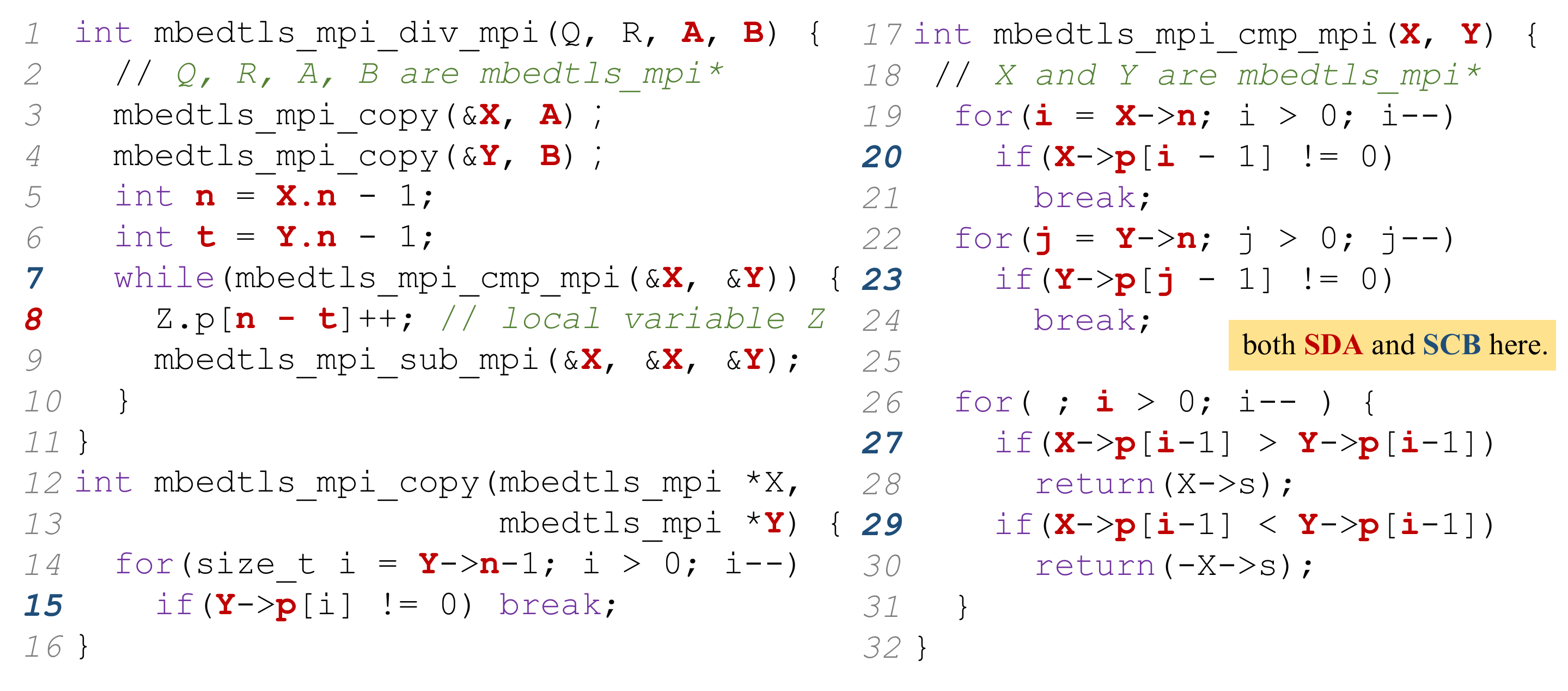}
  \vspace{-10pt}
  \caption{Vulnerable program points localized in MbedTLS 3.0.0. We mark the
  line numbers of \textcolor{pptred}{SDA} and \textcolor{pptgreen1}{SCB}.}
  \label{fig:mbedtls}
  \vspace{-5pt}
\end{figure*}

\smallskip
\parh{Case Study$_3$:}~\F~\ref{fig:mbedtls} shows leaking sites disclosed by \tool\ in MbedTLS 3.0.0.
In short, MbedTLS has similar implementation of BIGNUM with OpenSSL, where the
variable \Taint{n} in BIGNUM stores the number of leading zeros. Later computations
rely on \Taint{n} for optimization, for instance, the SDA at \Line{8}{fig:mbedtls}
in \texttt{mbedtls_mpi_div_mpi}@\Line{1}{fig:mbedtls}. It's worth noting that
\texttt{mbedtls_mpi_copy}@\Line{12}{fig:mbedtls}
is extensively called within
the life cycles of all involved BIGNUM, contributing to notable leaks in the
whole pipeline. Similar leaks also exist in MbedTLS 2.15.0. See our
website~\cite{snapshot} for more details.

\begin{figure}[!ht]
  % \vspace{-5pt}
  \centering
  \includegraphics[width=1.04\linewidth]{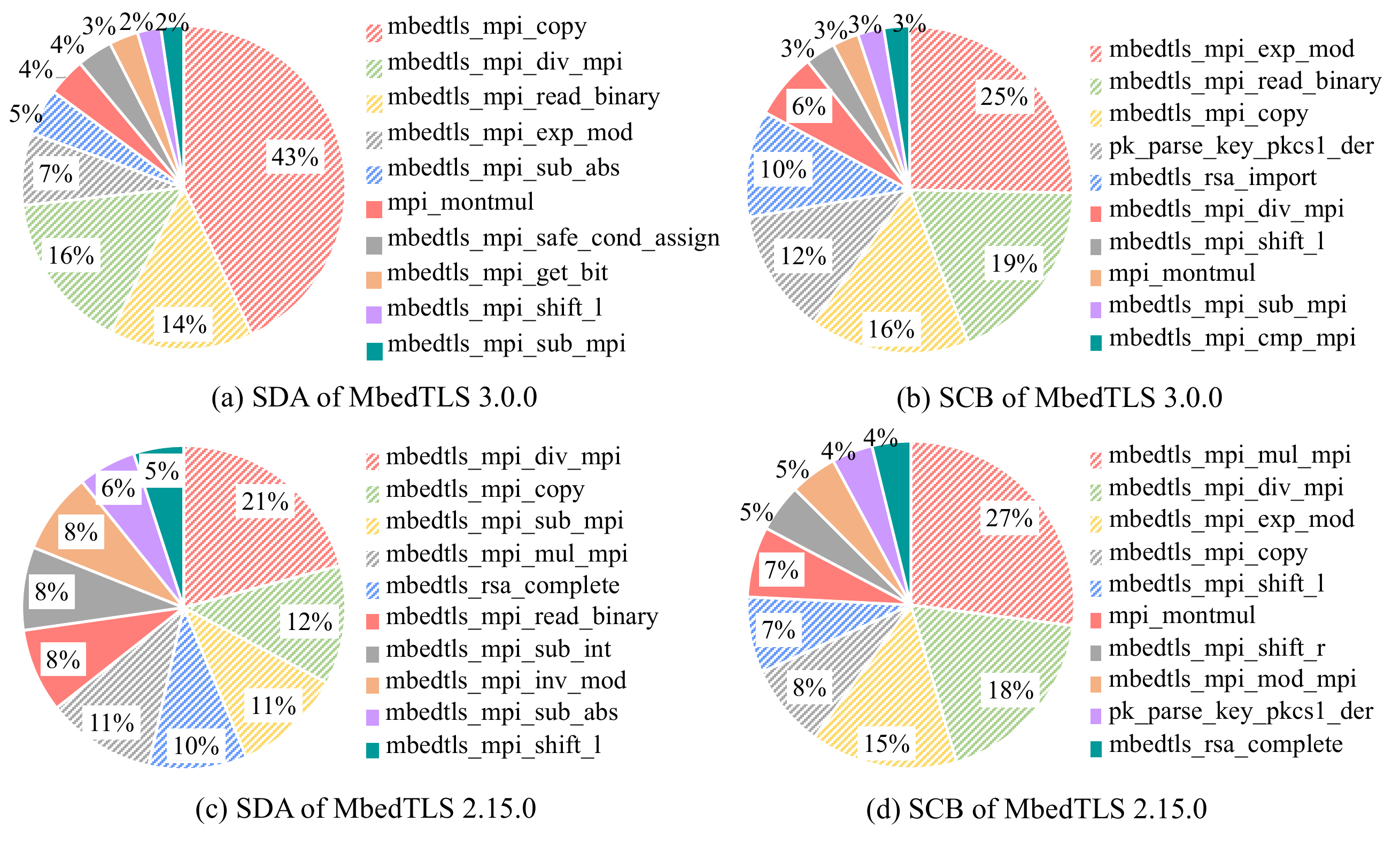}
  \vspace{-20pt}
  \caption{Distribution of top-10 functions in MbedTLS leaking most bits via
   either SDA or SCB vulnerabilities. Legend are in descending order.}
  \label{fig:pie-mbedtls}
  % \vspace{-10pt}
\end{figure}

\smallskip
\parh{Distribution.}~\F~\ref{fig:pie-mbedtls} reports the distribution of leaked bits among
top-10 vulnerable functions localized in MbedTLS. The two versions of MbedTLS
primarily leak bits in Pre-processing and have different
strategies when initializing BIGNUMs for CRT optimization. Thus, the
distributions of most vulnerable functions vary. For instance, the most
vulnerable functions in ver. 2.15.0 are for multiplication and division; they
are involved in calculating BIGNUMs for CRT. Notably, \texttt{mbedtls_mpi_copy}
is among the top-5 vulnerable functions on all four charts in
\F~\ref{fig:pie-mbedtls}. This function leaks the leading zeros of the input
BIGNUMs via both SDA and SCB. The \texttt{mbedtls_mpi_copy} function, as a
memory copy routine function, is frequently called (e.g., more than 1,000 times
in ver. 2.15.0). Though this function only leaks the leading zeros, given
that its input can be the private key or key-dependent intermediate value, the
accumulated leakage is substantial.

\begin{tcolorbox}[size=small]
  \textbf{Answer to RQ2}: \tool\ confirms all known flaws and identifies many
  new leakage sites, which span over the life cycle of cryptographic algorithms
  and exhibit diverse patterns. Distributions of leaked bits among vulnerable
  functions varies notably between software versions.
\end{tcolorbox}

\subsection{RQ3: Performance Comparison}
\label{subsec:eval-comparison}

To assess \tool's optimizations and re-formulations, we compare \tool\ with
previous tools on the speed, scalability, and capability of quantification and
localization.

\begin{table}[t]
  \caption{Padded length of side channel traces collected using Intel Pin.
  The above/below five rows are for SDA/SCB.}
  % \vspace{-5pt}
  \label{tab:trace-length}
  \centering
\resizebox{1.0\linewidth}{!}{
  \begin{tabular}{c|c|c|c|c}
    \hline
    & Pre-processing &     -      & Pre-processing & -          \\
    & Decryption     & Decryption & Decryption     & Decryption \\
    & Blinding       & Blinding   & -              & -          \\
    \hline
    OpenSSL 3.0.0 & $256 \times 256 \times 64$ & $256 \times 256 \times 15$ & $256 \times 256 \times 40$ & $256 \times 256 \times 9$\\ 
    % \hline
    OpenSSL 0.9.7 & $256 \times 256 \times 20$ & $256 \times 256 \times 14$ & $256 \times 256 \times 14$ & $256 \times 256 \times 13$ \\ 
    % \hline
    MbedTLS 3.0.0 & $256 \times 256 \times 16$ & $256 \times 256 \times 2$ & N/A & N/A\\ 
    % \hline
    MbedTLS 2.15.0 & $256 \times 256 \times 14$ & $256 \times 256 \times 2$ & N/A & N/A \\ 
    % \hline
    Libgcrypt 1.9.4 & $256 \times 256 \times 36$ & $256 \times 256 \times 22$ & $256 \times 256 \times 26$ & $256 \times 256 \times 25$ \\ 
    % \hline
    Libgcrypt 1.6.1 & $256 \times 256 \times 5$ & $128 \times 128 \times 19$ & $256 \times 256 \times 11$ & $256 \times 256 \times 10$ \\ 
    % \hline
    \hline
    OpenSSL 3.0.0 & $256 \times 256 \times 40$ & $256 \times 256 \times 10$ & $256 \times 256 \times 24$ & $256 \times 256 \times 4$ \\ 
    % \hline
    OpenSSL 0.9.7 & $256 \times 256 \times 8$ & $256 \times 256 \times 5$ & $256 \times 256 \times 5$ & $256 \times 256 \times 4$ \\ 
    % \hline
    MbedTLS 3.0.0 & $256 \times 256 \times 6$ & $256 \times 256 \times 1$ & N/A & N/A\\ 
    % \hline
    MbedTLS 2.15.0 & $256 \times 256 \times 6$ & $256 \times 256 \times 1$ & N//A & N/A \\ 
    % \hline
    Libgcrypt 1.9.4 & $256 \times 256 \times 12$ & $256 \times 256 \times 8$ & $256 \times 256 \times 10$ & $256 \times 256 \times 9$ \\ 
    % \hline
    Libgcrypt 1.6.1 & $256 \times 256 \times 3$ & $128 \times 128 \times 8$ & $256 \times 256 \times 6$ & $256 \times 256 \times 5$ \\ 
    \hline
  \end{tabular}
  }
  % \vspace{-10pt}
\end{table}

\smallskip
\parh{Trace Statistics.}~We report the lengths (after padding) of traces
collected using Pin in \T~\ref{tab:trace-length}. In short, all traces collected
from real-world cryptosystems are lengthy, imposing high challenge for analysis.
Nevertheless, \tool employs encoding module $\mathcal{S}$ and compressing module
$\mathcal{R}$ to effectively process lengthy and sparse traces, as noted in
\S~\ref{sec:framework}.

\smallskip
\parh{Impact of Re-Formulations/Optimizations.}~\tool\ casts MI as CP when
quantifying the leaks. This re-formulation is faster (see comparison below) and
more precise, because calculating MI via MP (as done in MicroWalk) cannot
distinguish blinding in traces. As reported in \T~\ref{tab:trace-length}, a
great number of records are related to blinding, and they lead to false
positives of MicroWalk. For localization, the unoptimized Shapley value has
$\mathcal{O}(2^N)$ computing cost. Given the trace length $N$ is often extremely
large (\T~\ref{tab:trace-length}), computing Shapley value is infeasible. With
our domain-specific optimizations, the cost is reduced as nearly constant.

\subsubsection{Time Cost and Scalability}
\label{subsubsec:scalability}

\begin{table}[t]
  \caption{Scalability comparison of static- or trace-based tools.}
  % \vspace{-5pt}
  \label{tab:scalability}
  \centering
\resizebox{0.95\linewidth}{!}{
  \begin{tabular}{c|c|c|c|c}
    \hline
               & CacheD & Abacus & CacheS & CacheAudit \\
    \hline
    Technique  & \multicolumn{2}{c|}{symbolic execution} & \multicolumn{2}{c}{abstract interpretation} \\
    \hline
    Libgcrypt  & fail ($>$ 48h) & fail ($>$ 48h) & fail & fail  \\
    \hline
    Libjpeg    & fail   & fail   & fail & fail       \\
    \hline
  \end{tabular}
  }
  \vspace{-5pt}
\end{table}

\parh{Scalability Issue of Static-/Trace-Based Tools.}~As noted in \Circ{7}
in \S~\ref{sec:requirement}, prior static- or trace-based analyses rely on
expensive and less scalable techniques. They, by default, primarily analyze a
program/trace cut and neglect those pre-processing functions in cryptographic
libraries. To faithfully assess their capabilities, we configure them to analyze
the entire trace/software (which needs some tweaks on their codebase). We
benchmark them on Libjpeg and RSA of Libgcrypt 1.9.4.
Abacus/CacheD/CacheS/CacheAudit can only analyze 32-bit x86 executable. We thus
compile 32-bit Libgcrypt and Libjpeg.
Results are in \T~\ref{tab:scalability}. CacheS and CacheAudit throw exceptions
of unhandled x86 instruction. Both tools, using rigorous albeit expensive
abstraction interpretation, appear to handle a subset of x86 instructions.
Fixing each unhandled instruction would presumably require defining a new
abstract operator~\cite{cousot1977abstract}, which is challenging on our end.
Abacus and CacheD can be configured to analyze the full trace of Libgcrypt.
Nevertheless, both of them fail (in front of unhandled x86 instructions) after
about 48h of processing. In contrast, \tool\ takes less than 17h to finish the
training and analysis of the Libcrypt case; see \T~\ref{tab:time}.

\begin{table}[t]
  \caption{Training time of 50 epochs for the RSA cases.}
  \label{tab:time}
  \centering
\resizebox{1.0\linewidth}{!}{
  \begin{tabular}{c|cccc|cccc}
    \hline
    % \multirow{3}{*}{\shortstack{Pre-processing\\Decryption\\Blinding}}
    & \multicolumn{4}{c|}{SDA} & \multicolumn{4}{c}{SCB} \\
    \hline
    \multirow{3}{*}{Configuration}  & Pre.   &  -     & Pre. & -    & Pre.   & -      & Pre. & -      \\
                                    & Dec.   & Dec.   & Dec. & Dec. & Dec.   & Dec.   & Dec. & Dec.   \\
                                    & Blind. & Blind. & -    & -    & Blind. & Blind. & -    & -      \\
    \hline
    OpenSSL 3.0.0 & $22$h & $5$h & $3$h & $50$min & $13$h & $3$h & $2.5$h & $20$min \\ 
    \hline
    OpenSSL 0.9.7 & $6.5$h & $5$h & $1$h & $1$h & $2.5$h & $1.5$h & $25$min & $20$min \\ 
    \hline
    MbedTLS 3.0.0 & $5$h & $40$min & N/A & N/A & $2.5$h & $20$min & N/A & N/A \\ 
    \hline
    MbedTLS 2.15.0 & $5$h & $40$min & N/A & N/A & $2.5$h & $20$min & N/A & N/A \\ 
    \hline
    Libgcrypt 1.9.4 & $12.5$h & $7.5$h & $1.5$h & $1.5$h & $4$h & $2.5$h & $50$min & $45$min \\ 
    \hline
    Libgcrypt 1.6.1 & $1.5$h & $1.5$h & $55$min & $50$min & $1$h & $40$min & $30$min & $25$min \\ 
    \hline
  \end{tabular}
  }
  \begin{tablenotes}
    \footnotesize
    \item 1. Due to the limited space, we use Pre., Dec., and Blind. to denote
    Pre-processing, Decryption, and Blinding, respectively.
    \item 2. Blind. has $\times 4$ training samples.
  \end{tablenotes}
  \vspace{-15pt}
\end{table}

\smallskip
\parh{Training/Analyzing Time of \tool.}~\T~\ref{tab:time} presents the RSA case training time, which is
calculated over 50 epochs (the maximal epochs required) on one Nvidia GeForce
RTX 2080 GPU. In practice, most cases can finish in less than 50 epochs.
For AES-128, training 50 epochs takes about 2 mins. Training 50 epochs for
Libjpeg/PathOHeap takes 2-3 hours. As discussed in \S~\ref{subsec:pdf-est},
since we transform computing MI as estimating CP, \tool\ only needs to be
trained (for estimating CP) once. Once trained, it can analyze 256 traces in
\textbf{1-2 seconds} on one Nvidia GeForce RTX 2080 GPU, and less than
\textbf{20 seconds} on Intel Xeon CPU E5-2683 of \textbf{4} cores.

In sum, \tool\ is much faster than existing trace-based/static tools. By using
CP, it principally reduces computing cost comparing with conventional dynamic
tools (see \S~\ref{sec:mi}). We also note that it is hard to make a fully fair
comparison: training \tool\ can use GPU while existing tools \textit{only}
support to use CPUs. Though \tool\ has smaller time cost on the GPU (Nvidia 2080
is \textit{not} very powerful), we do not claim \tool\ is faster than prior
dynamic tools. In contrast, we only aim to justify that \tool\ is \textit{not}
as heavyweight as audiences may expect. Enabled by our theoretical and
implementation-wise optimizations, \tool\ efficiently analyzes complex
production software.

\subsubsection{Capability of Quantification and Localization}
\label{subsubsec:small}

\parh{Small Programs and Trace cuts.}~As evaluated in
\S~\ref{subsubsec:scalability}, previous static-/trace-based tools are
incapable of analyzing the full side channel traces. Therefore, we compare
them with \tool\ using small program (e.g., AES) and trace cuts.

Overall, the speed of \tool\ (i.e., training + analyzing) still largely
outperforms static/trace-based methods. For instance, CacheD~\cite{wang2017cached},
a qualitative tool using symbolic execution, takes about 3.2 hours to analyze
only the decryption routine of RSA in Libgcrypt 1.6.1 without considering blinding.
\tool\ takes under one hour for this setting. In addition,
Abacus~\cite{bao2021abacus}, which performs quantitative analysis with
symbolic execution, requires 109 hours to process one trace of Libgcrypt 1.8.
Note that it only analyzes the decryption module (several caller/callee
functions) without considering the blinding, pre-rocessing functions, etc. In
contrast, \tool\ can finish the training within 2 hours (the trace length of
Libgcrypt 1.9 is about the same as ver. 1.8) in this setting. It's worth noting
that, \tool\ only needs to be trained for once, and it takes only several seconds
to analyze one trace. That is, when analyzing multiple traces, previous tools
has fold increase on the time cost whereas \tool\ only adds several seconds.

The quantification/localization precision of \tool\ is also much higher.
Abacus reports 413.6 bits of leakage for AES-128 (it neglects dependency among
leakage sites, such that the same secret bits can be repetitively counted at
different leakage sites), which is an overestimation since the key has only 128
bits. For RSA trace cuts, Abacus under-quantifies the leaked bits because it
misses many vulnerabilities due to implicit information-flow. When localizing
vulnerabilities in AES-128, we note that all static/trace-based have correct
results. For localization results of RSA trace cuts,
see~\S~\ref{subsubsec:categorization}. In short, none of the previous tools
can identify all the categories of vulnerabilities.

\smallskip
\parh{Dynamic Tools.}~Previous dynamic tools do not suffer from the scalability
issue and have comparable speed with \tool. Nevertheless, they require
re-launching their whole pipeline (e.g., sampling + analyzing) for each trace.

For quantification, MicroWalk over-quantifies the leaked bits of RSA as 1024
when blinding is enabled, since it regards random records as vulnerable.
Similarly, it reports that ORAM cases have all bits leaked despite they are
indeed secure. MicroWalk can correctly quantify the leaks of whole traces
for AES cases and constant-time implementations, because no randomness exists.
Nevertheless, since the same key bits are repeatedly reused on the trace,
its quantification results for single record, when summed up, are incorrectly
inconsistent with the result of whole trace.
For localization, as summarized in \S~\ref{subsubsec:categorization}, previous
dynamic tools are also either incapable of identifying all categories of
vulnerabilities or yields many false positive. For instance, MicroWalk can
regards all records related to blinding (over 1M in OpenSSL 3.0.0; see trace
statistics in \T~\ref{tab:trace-length}) as ``vulnerable''.

\begin{tcolorbox}[size=small]
  \textbf{Answer to RQ3}: With domain-specific transformations and optimizations
  applied, \tool\ addresses inherent challenges like non-determinism, and
  features fast, scalable, and precise quantification/localization. Evaluations
  show its advantage over previous tools.
\end{tcolorbox}
\vspace{-5pt}

% \vspace{-5pt}
\section{Discussion}
\label{sec:discussion}
% \vspace{-5pt}

\parh{Handling Real-World Attack Logs.}~Side channel observations (e.g.,
obtained in cross-virtual machine attacks~\cite{Zhang12}), are typically noisy.
\tool\ handles real attack logs by considering noise as non-determinism (see
\S~\ref{subsec:non-det}), thus quantifying leaked bits in those logs.
Nevertheless, we do not recommend localizing vulnerabilities using real attack
logs, since mapping these records back to program statements are challenging.
Pin is sufficient for developers to ``debug''.

\smallskip
\parh{Analyzing Media Data.}~\tool\ can smoothly quantify and localize
information leaks for media software. Unlike previous static-/trace-based tools,
which require re-implementing the pipeline to model floating-point instructions
for symbolic execution or abstract interpretation, \tool\ only needs the
compressor $\mathcal{R}$ to be changed. In addition, \tool\ is based on NN,
which facilitates extracting ``contents'' of media data to quantify leaks,
rather than simply comparing data byte differences. See evaluation results in
\Appx~\ref{appx:r-detail}.

\smallskip
\parh{Program Taking Public Inputs.}~We deem that different public inputs should
not largely influence our analysis over cryptographic and media libraries,
whose reasons are two-fold. First, for cryptosystems like OpenSSL, the public
inputs (i.e., plaintext or ciphertext) has a relatively minor impact on the
program execution flow. To our observation, public input values only influence a
few loop iterations and \texttt{if} conditions. Media libraries mainly process
private user inputs, which has no ``public inputs''. In practice, the influences
of public inputs (including other non-secret local variables) are treated as
non-determinism by \tool. That is, they are handled consistently as how \tool\
handles cryptographic blinding (\S~\ref{subsec:non-det}), because neither is related
to secret.

Given that said, configurations (e.g., cryptographic algorithm or image compression
mode) may notably change the execution and the logged execution traces. We view
that as an orthogonal factor. Moreover, modes of cryptographic algorithms and media
processing procedures are limited. Users of \tool\ are suggested to fix the mode
before launching an analysis with \tool, then use another mode, and so on.

\smallskip
\parh{Keystroke Templating Attacks.}~Quantifying and localizing the information
leaks that enable keystroke templating attacks should be feasible to explore.
With side channel traces logged by Intel Pin, machine learning is used to
predict the user's key press~\cite{wang2019unveiling}. Given sufficient data
logged in this scenario, \tool\ can be directly applied to quantify the leaked
information and localize leakage sites.

\smallskip
\parh{Large Software Monoliths.}~For analyzing complex software like browsers
and office products, our experience is that using Intel Pin to perform dynamic
instrumentation for production browsers is difficult. With this regard, we
anticipate adopting other dynamic instrumentors, if possible, to enable
localizing leaks in these software. With correctly logged execution trace,
\tool\ can quantify the leaked bits and attribute the bits to side channel
records. 

\smallskip
\parh{Training Dataset Generalization.}~One may question to what degree traces
obtained from one program can be used as training set for detecting leaks in
another. In our current setup, we do not advocate such a ``transfer'' setting.
Holistically, \tool\ learns to compute PD by accurately distinguishing traces
produced when the software is processing different secret inputs. By learning
such distinguishability, \tool\ eliminates the need for users to label the
leakage bits of each training data sample. Nevertheless, knowledge learned for
distinguishability may differ between programs. It is intriguing to explore
training a ``general'' model that can quantify different side channel logs,
particularly when collecting traces for the target program is costly. To do so,
we expect to incorporate advanced training techniques (such as transfer
learning~\cite{pan2009survey}) into our pipeline.

\smallskip
\parh{Key Generation.}~In RSA evaluations, we feed cryptographic libraries with the key
in a file. That is, the cryptographic library execution does not involve ``key
generation'', which is \textit{not} due to ``limited coverage'' of \tool.
Previous works~\cite{moghimi2020copycat} have exploited RSA key generation.
With manual effort, we find that the key generation functions heavily use BIGNUM,
involving vulnerable BIGNUM initialization and computation functions already
localized by \tool\ in \S~\ref{subsec:eval-localization}, e.g.,
\texttt{BN_bin2bn} in \F~\ref{fig:openssl} and \texttt{BN_sub} (see our full
report~\cite{snapshot}). 

\smallskip
\parh{BIGNUM Implementation.}~We also investigate other cryptosystems.
LibreSSL and BoringSSL are built on OpenSSL. Their BIGNUM
implementations and OpenSSL share similar vulnerable coding patterns (i.e., the
leading zero leak patterns; see \Type{C} of \S~\ref{subsec:eval-localization}).
We also find similar BIGNUM vulnerable patterns in Botan (see~\cite{botan-bn}).
In contrast, we find Intel IPP does not use an individual variable to
record \#leading zeros in BIGNUM (see~\cite{intel-ipp-bn}), hence it
is likely free of \Type{C}.

% \vspace{-5pt}
\section{Conclusion}
\label{sec:conclusion}
% \vspace{-5pt}

We present \tool\ to quantify cache side channel leakages via MI. We also
formulate secret leak as a cooperative game and enable localization via
Shapley value. Our evaluation shows that \tool\ overcomes typical hurdles (e.g.,
scalability, accuracy) of prior works, and computes information leaks in
real-world cryptographic and media software.

\section*{Acknowledgement}
% \vspace{-5pt}

We thank all anonymous reviewers and our shepherd for their valuable feedback.
We also thank Janos Follath, Matt Caswell, and developers of Libjpeg-turbo
for their prompt responses and comments on our reported vulnerabilities.

\bibliographystyle{plain}
\bibliography{bib/main}

\begin{appendix}

\section{Libjpeg and \pp\ Evaluation}
\label{appx:extended-eval}

\parh{Data Preparing.}~We choose the CelebA dataset~\cite{liu2015faceattributes}
as the image inputs of Libjpeg. The CelebA consists of 160,000 training and 10,000
validation images. We use the training images and their corresponding side channel
traces to estimate CP via \model. The validation images and their induced side
channels are adopted for de-biasing (in case there exists non-determinism).

\parh{Trace Logging.}~Following the same configuration of
\S~\ref{sec:evaluation}, we use Pin to log execution traces of Libjpeg.

\parh{Logging via \pp.}~Besides using Pin, we collect cache set access traces
(for both cryptosystems and Libjpeg) via
\pp~\cite{tromer2010efficient}, in userspace-only scenarios.
Following~\cite{yuan2022automated}, we use Mastik~\cite{yarom2016mastik}, a
micro-architectural side channel toolkit, to perform \pp\ and log victim's
access toward L1D and L1I cache. We use Linux \texttt{taskset} to pin victim
software and the spy process on the same CPU core. Scripts of \pp\
experiments are at~\cite{snapshot}.

\subsection{Libjpeg}
\label{appx:libjpeg}

\parh{Quantification.}~Libjpeg does not contain mitigations like blinding. Thus, side channels
collected by Pin are deterministic. Nevertheless, we argue that merely considering
the difference among SDA/SCB, which is a conventional setup for cryptosystems,
will \textit{over-estimate} the leakage. The reasons are two-fold: 1) Traces can
be largely divergent by tweaking trivial pixels in the input images, e.g.,
background color pixels. 2) Even if all pixels are equally sensitive, the
leakage may be insufficient to recover input images. For instance, attackers
only infer whether each pixel value is larger than a threshold (as how keys are
usually recovered) and obtain a binary image --- it's still infeasible to
recognize humans in such images.

To handle these, we extend the \textit{generalizability} consideration, which is
proposed to handle non-deterministic side channels derived from cryptosystems, to quantify deterministic side channels made by Libjpeg. Therefore,
our quantification is the same as for processing non-deterministic side channels
of cryptosystems. Due to the aforementioned issue, we deem trivial pixels
contain little information. As introduced in \S~\ref{subsec:non-det}, \tool\
should accordingly report a zero leakage since these non-sensitive factors are
not generalizable.

Given that the \#images is infinite, it's infeasible to decide the value of
$p(F)/p(T)$ in \E~\ref{equ:cond-prob} which should be a constant. We thus report
the leakage ratio for Libjpeg. As shown in \T~\ref{tab:quant-libjpeg}, leakage
ratios reach 100\% for SDA and SCB if we only consider the
\textit{distinguishability}, which indicates side channels can differ images.
Nevertheless, the leakage ratios are reduced to 93\% for SDA and 49\% for SCB
when we faithfully take \textit{generalizability} into account. To further prove
that \tool\ indeed captures the critical information in face photos, we present
the output of $\mathcal{R}$ in \F~\ref{fig:face}. The output manifests
common human face features like eyes and noses. 

\begin{table}[t]
  \caption{Leakage ratios (see \E~\ref{equ:non-deter}) of Libjpeg
  without $\rightarrow$ with considering generalizability.}
  \label{tab:quant-libjpeg}
  % \vspace{-5pt}
  \centering
\resizebox{0.7\linewidth}{!}{
  \begin{tabular}{c|c|c}
    \hline
               & SDA     & SCB \\
    \hline
    cache line & 100\% $\rightarrow$ 93\%   & 100\% $\rightarrow$ 49\%  \\
    \hline
  \end{tabular}
  }
% \vspace{-5pt}
\end{table}

\begin{table}[t]
  \caption{Representative vulnerable functions
  localized in Libjpeg and their types.}
  \label{tab:loc-libjpeg}
  % \vspace{-5pt}
  \centering
\resizebox{0.75\linewidth}{!}{
  \begin{tabular}{c|c}
    \hline
    Function & Type \\
    \hline
    \texttt{decode_mcu} & SDA, SCB \\
    \hline
    \texttt{jsimd_ycc_extbgrx_convert_avx2} & SDA \\
    \hline
    \texttt{jsimd_idct_islow_avx2} & SDA, SCB \\
    \hline
  \end{tabular}
  }
% \vspace{-5pt}
\end{table}

\parh{Localization.}~Representative vulnerable functions of Libjpeg are given in
\T~\ref{tab:loc-libjpeg}. The leaked bits are spread across hundreds of program
points, mostly from the IDCT, YCC encoding, and MCU modules. Libjpeg converts
JPEG images into bitmaps, whose procedure has many SDA and SCB. We manually
checked our findings, which are \textit{aligned} with~\cite{yuan2022automated}.
Nevertheless, the YCC encoding-related functions are newly found by us. \tool\
also shows that SDA in IDCT leaks the most bits, whereas the MCU modules leak
more bits via SCB.~\cite{yuan2022automated} flags those issues without
quantification.

\begin{table}[t]
  \caption{Leaks of side channels collected via \pp.}
  % \vspace{-5pt}
  \label{tab:quant-pp}
  \centering
\resizebox{1.0\linewidth}{!}{
  \begin{tabular}{c|c|c|c|c}
    \hline
                & RSA D & RSA D & RSA D & RSA D \\
    \hline
    \#Repeating & 1 & 2 & 4 & 8 \\
    \hline
    Leakage     & 19.3 (1.8\%) & 29.4 (2.8\%) & 34.9 (3.4\%) & 35.5 (3.4\%)  \\
    \hline
                & RSA D & RSA I & Libjpeg D & Libjpeg I\\
    \hline
    \#Repeating & 16 & 8 & 8 & 8 \\
    \hline
    Leakage     & 35.6 (3.4\%)  & 21.6 (2.1\%) & 20.8\% & 12.9\% \\
    \hline
  \end{tabular}
  }
  \begin{tablenotes}
    \footnotesize
    \item 1. ``D'' denotes L1 D cache whereas ``I'' denotes L1 I cache.
    \item 2. We also report leakage ratios for RSA cases to ease comparison.
  \end{tablenotes}
% \vspace{-15pt}
\end{table}

\subsection{\pp}
\label{appx:pp}

Following previous setups~\cite{yuan2022automated}, a common \pp\ is launched on
the same core with the victim software and periodically probes L1 cache when the
victim software is executed. This mimics a practical and powerful attacker and
is also the default setup of the \pp\ toolkit~\cite{yarom2016mastik} leveraged
in our evaluation and relevant research in this field.
To prepare traces about Pre-processing, we halt the victim program after
the pre-processing stage.
Generally, launching \pp\ is costly. Without loss of generality, we use \pp\ to collect RSA
side channels from OpenSSL 0.9.7 and Libjpeg. Attackers often repeat launching
\pp~\cite{Zhang12}. As in \T~\ref{tab:quant-pp}, repeating an attack to collect
more logs does improve information, but only marginal. Compared to merely doing
\pp\ once, repeating $\times$4 yields more information. However, repeating more
times does not necessarily improve, since the information source is always the
same.
We also note that Pre-processing has more leaks: for two (RSA, \#Repeating=8) cases of
L1 D and I cache (i.e., the 2nd and 3rd columns in the last row), Pre-processing has
22.5 (total 35.5) and 14.1 (total 21.6) leaked bits.

Libjpeg leaks more information than RSA. Not like recovering keys where each key
bit needs to be analyzed, recovering every pixel is not necessary for inferring
images. As previously stated~\cite{yuan2022automated}, pixels conform to specific
constraints to form meaningful contents (e.g., a human face), which typically
have lower dimensions than pixel values. As a result, extracting these
constraints can give rich information already.

\section{Transformations in $\mathcal{R}$}
\label{appx:r-detail}

As stated in \S~\ref{sec:framework}, we adopt different mathematical
transformations in $\mathcal{R}$ for media data and cryptographic keys.
In what follows, we elaborate on compressing latent vectors of
media data and secret keys with different transformations in $\mathcal{R}$.

\parh{Media Data.}~We use image to demonstrate the cases for
media data; the conclusion can be extended to other media data
straightforward~\cite{yuan2022automated}. It is generally obscure to measure the
amount of information encoded in high-dimensional media data like
images~\cite{zhang2018unreasonable}. To extract information, an image, before
being fed into modern neural networks (e.g., classifier $\mathcal{C}$ in \tool),
is generally normalized into $[-1, 1]$ and represented as a
$channel \times width \times height$ matrix~\cite{krizhevsky2012imagenet}. Let
a private image be $k_m$ and the associated side channel log be $o_m$, we set
the output of $\mathcal{S}(o_{m})$ as a matrix of the same size. Accordingly,
our compressor $\mathcal{R}_{m}$ is implemented using the
$tanh: [-\infty, \infty] \rightarrow [-1, 1]$ function, facilitating
$H(o_m) \leq H(k_m)$. $tanh$ is parameter free
(i.e., $\theta_{\mathcal{R}} = \emptyset$), eliminating extra training cost.
Also, if certain properties in an image are particularly desirable by
attackers, e.g., the gender of portrait images, $k_m$ and the matrix from
$\mathcal{S}(o_{m})$ can be replaced with a vectorized representation for image
properties. We refer readers to~\cite{pumarola2018ganimation} for
vectorizing image attributes.

\parh{Cryptographic Keys.}~For a cryptographic key $k_c$ of length $L$, there
are total $2^{L}$ uniformly distributed key instances. The information in one
key instance is thus $H(k) = \log 2^{L} = L$ bits. Since a key only contains
binary values and neural networks are hard to be deployed with binary
parameters, for the associated side channel log $o_c$,
we need to transform the floating-point encoding output $\mathcal{S}(o_c)$,
which is accordingly a vector of length $L$, into binary bits. Existing
profiling-based side channel attacks map side channels to keys via $L$ bit-wise
classification tasks ~\cite{hospodar2011machine,kim2019make,hettwer2018profiled}.
In each task, a floating point is transformed to 1 if it is greater than a
threshold. Nevertheless, this transformation is not applicable in $\mathcal{R}$,
given that it is not differentiable, which impedes the optimization of $\theta$.
Inspired by the optimization for binary variables
~\cite{courbariaux2015binaryconnect,courbariaux2016binarized}, we design
$\mathcal{R}_{c}$ by joining two components: 1) a non-parametric $sigmoid:
[-\infty, \infty] \rightarrow [0, 1]$ function which generates $L$ independent
(as bits in $k_c$ are independent) probabilities, and 2) a parametric Bernoulli
distribution $Bern: \Pr(Bern(\cdot)=1) + \Pr(Bern(\cdot)=0) = 1$ which takes
its inputs, i.e., $sigmoid(\mathcal{S}(o_{c}))$, as parameters for optimization.
Thus, we have $\mathcal{R}_{c} = Bern \circ sigmoid$ and
$\theta_{\mathcal{R}_{c}} = sigmoid(\mathcal{S}(o_{c}))$.
See our codebase~\cite{snapshot} for details.

\begin{figure}[!ht]
    % \hspace{-5pt}
    \centering
    \includegraphics[width=0.8\linewidth]{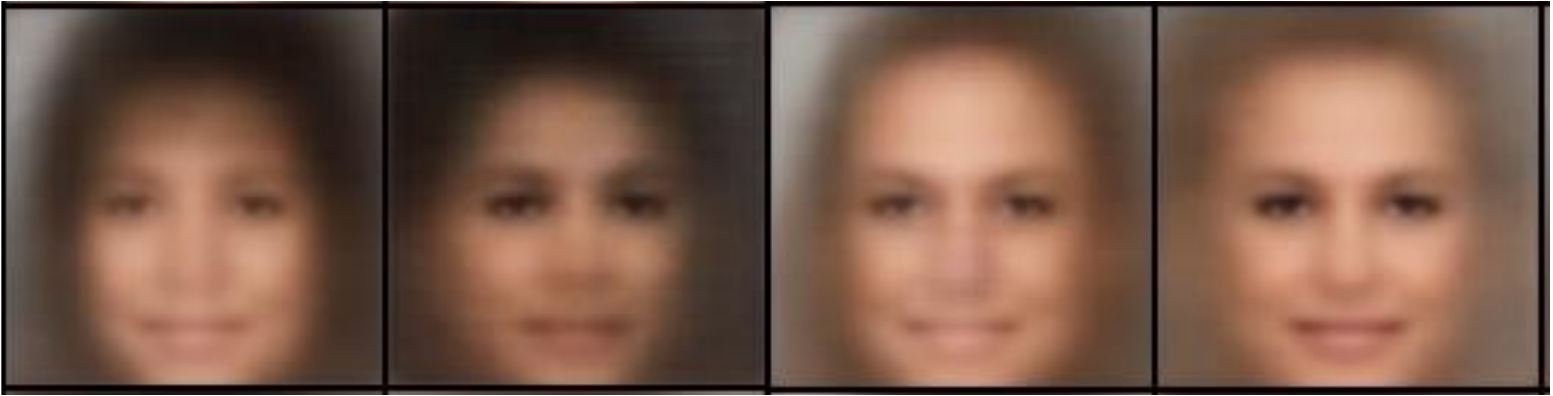}
    \vspace{-5pt}
    \caption{Outputs of $\mathcal{R}$ when quantifying leakage of images.}
    % \vspace{-10pt}
    \label{fig:face}
\end{figure}

Recall as we discussed in \S~\ref{sec:framework}, $\mathcal{R}$ is designed for
shrinking maximal information in a side channel trace $o$, enabled by this
principle, \tool\ is ``forced'' to focus on secret-related information. 
Outputs of $\mathcal{R}$ are shown in \F~\ref{fig:face}: it captures common
facial features, such as eyes and mouth, when quantifying information leaks
for human photos.

\section{Reconstruct Secrets From Side Channels}
\label{appx:reconstruction}

\Appx~\ref{appx:r-detail} demonstrates that $\mathcal{R}$ captures facial attributes
when quantifying leakage of portrait photos. Readers may question whether \tool\
can directly reconstruct secrets from side channels.
To this end, we extend \tool\ by tweaking the optimization objectives (only adds
\textit{one} line of code; see our manuals~\cite{snapshot}), and present results in
\T~\ref{tab:recons}. We show reconstructed images on our
website~\cite{snapshot}. Overall, keys/images with reasonable quality are reconstructed using traces logged by
Intel Pin or \pp. In particular, reconstructed images have more vivid appearance
than \F~\ref{fig:face}; it's reasonable since quantifying leaks emphasizes on
the \textit{distinguishability}. These promising results illustrate the
superiority of \tool's design and extendability, and the exploitability of
localized program points. Having stated that, we clarify that aligned with prior
works in this field, \tool\ is a bug detector, \textit{not} for reconstructing
secret in real attacks. Also, as a proof of concept, the extended \tool\ shows
the feasibility of key/image reconstruction; the quality of reconstructed
keys/images may be further improved~\cite{yuan2022automated}, which we leave as
one future work.

\begin{table}[t]
    \caption{Quantitative evaluation of the reconstruction.}
    % \vspace{-5pt}
    \label{tab:recons}
    \centering
  \resizebox{1.0\linewidth}{!}{
    \begin{tabular}{c|c|c|c}
      \hline
                 & $^{*}$Libjpeg (Pin) & $^{\dagger}$OpenSSL 0.9.7 (Pin) & $^{*}$Libjpeg (Prime+Probe) \\
      \hline
      DA/D cache & 41.5\%             & 58.2\%              & 36.8\% \\
      \hline
      CB/I cache & 40.4\%         & 57.7\%              & 36.0\% \\
      \hline
    \end{tabular}
    }
    \begin{tablenotes}
      \footnotesize
      \item * We report percentage of reconstructed photos that are recognized as the
      same faces with the reference photos. We use the criterion of~\cite{yuan2022automated}.
      \item $\dagger$ Pre-processing + Decryption with blinding enabled.
      We report percentage of correct bits. Random guess (baseline) should be 50\%.
    \end{tablenotes}
    % \vspace{-10pt}
  \end{table}
\section{Extended Proof of \E~\ref{equ:appx}}
\label{appx:correctness}

This Appendix section proves \E~\ref{equ:appx} in a two-step approach; we refer
the following proof skeleton to~\cite{belghazi2018mutual,tsai2020neural}.
Since $c(k, o)$ is only decided by $\mathcal{F}_{\theta}$, for simplicity, we also
denote it as a parameterized function. 
In particular, Lemma~\ref{lem:est} first shows that the empirically measured
$\hat{I}^{(n)}_{\theta^{\dagger}}(K; O)$ is consistent with
$\hat{I}_{\theta^{\dagger}}(K; O)$. Lemma~\ref{lem:appx} then proves that
$\hat{I}_{\theta^{\dagger}}(K; O)$ stays close to $I(K; O)$.
In all, Tsai et al.~\cite{tsai2020neural} has pointed out that the following two
prepositions hold for neural networks:

\begin{proposition}[Boundness]
    \label{prop:boundness}
    $\forall \hat{c}_{\theta}$ and $\forall k, o$, there exist two constant
    bounds $B_l$ and $B_u$ such that $B_l \leq \log \hat{c}_{\theta}(k, o) \leq B_u$.
\end{proposition}

\begin{proposition}[$\log$-smoothness]
    \label{prop:log}
    $\forall k, o$ and $\forall \theta_1, \theta_2 \in \Theta$, 
    $\exists \alpha > 0$ such that
    $| \log\hat{c}_{\theta_1}(k, o) - \log\hat{c}_{\theta_2}(k, o)| \leq \alpha \| \theta_1 - \theta_2 \|$.
\end{proposition}

\Prop~\ref{prop:boundness} states that outputs of a neural network are bounded
and \Prop~\ref{prop:log} clarifies that the output of a neural network will not
change too much if the parameter $\theta$ change slightly~\cite{tsai2020neural}.
By incorporating the bounded rate of uniform convergence on parameterized
functions~\cite{bartlett1998sample}, we have:

\begin{lemma}[Estimation]
    \label{lem:est}
    $\forall \epsilon > 0$,
    \begin{equation*}
        \begin{aligned}
            &\Pr\limits_{\{\langle k, o \rangle \}^{(n)}} \left( 
                \sup\limits_{\hat{c}_{\theta^{\dagger}} \in \mathcal{C}} \left| 
                    \hat{I}_{\theta^{\dagger}}^{(n)}(K; O) - \mathbb{E}_{P_{K \times O}} [
                        \log \hat{c}_{\theta^{\dagger}} (k, o)
                    ]
                \right| \geq \epsilon
            \right) \\
            &\leq 2 |\Theta| \exp \left(
                \frac{- n \epsilon^2}{2(B_u - B_l)^2}
            \right).
            % = O\left( \frac{1}{e^n} \right).
        \end{aligned}
    \end{equation*}
\end{lemma}

%\vspace{4cm}
\Lem~\ref{lem:est} applies the classical consistency
theorem~\cite{geer2000empirical} for extremum estimators. Here, extremum
estimators denote parametric functions optimized via maximizing or minimizing
certain objectives; note that $\hat{c}_\theta$ is optimized to maximize the
binary cross-entropy in \E~\ref{equ:bce}. It illustrates that
$\hat{I}^{(n)}_{\theta^{\dagger}}(K; O)$ convergent to
$\hat{I}_{\theta^{\dagger}}(K; O)$ as $n$ grows. Future, based the universal
approximation theory of neural networks~\cite{hornik1989multilayer}, we have:

\begin{lemma}[Approximation]
    \label{lem:appx}
    $\forall \epsilon > 0$, there exists a family of neural networks
    $N = \{ \hat{c}_{\theta}: \theta \in \Theta\}$ such that
    \begin{equation*}
    \inf\limits_{\hat{c}_{\theta} \in N}
    |\mathbb{E}_{P_{K \times O}}[\log \hat{c}_{\theta}(k, o)] - I(K, O)|
    \leq \epsilon.
    \end{equation*}
\end{lemma}

\Lem~\ref{lem:appx} states that $\hat{I}_{\theta^{\dagger}}(K; O)$ can
approximate $I(K; O)$ with arbitrary accuracy. Therefore, \E~\ref{equ:appx} can
be derived from \Lem~\ref{lem:est} and \Lem~\ref{lem:appx} based on the
triangular inequality. 
\section{Key Properties of Shapley Values}
\label{appx:sv-properties}

Following \S~\ref{sec:local}, this Appendix shows several key properties of
Shapley value. We discuss why it is suitable for side channel analysis and how
the properties help our localization.
To clarify, the following theorems are \textit{not} proposed by this paper;
they are for reader's reference only.

\vspace{-5pt}
\begin{theorem}[Efficiency~\cite{shapley201617}]
    \label{th:efficiency}
    The sum of Shapley value of all participants (i.e., side channel records)
    equals to the value of the grand coalition:
    \begin{equation}
        \sum_{i \in R^o} \pi_{i}(\phi) = \phi(o) - \phi(o_{\emptyset}).
    \end{equation}
\end{theorem}

\Th~\ref{th:efficiency} states that the assigned Shapley value for each side
channel record satisfies the apportionment defined in \Df~\ref{def:apportion}.
$\phi(o_{\emptyset}) = 0$ since an empty $o$ leaks no secret.

\vspace{-5pt}
\begin{theorem}[Symmetry~\cite{shapley201617}]
    \label{th:symmetry}
    If $\forall S \subseteq R^{o} \setminus \{i,j\}$, $i$- and $j$-th players
    are equivalent, i.e., $\phi(o_{S \cup \{i\}}) = \phi(o_{S \cup \{j\}})$,
    then $\pi_{i}(\phi) = \pi_{j}(\phi)$.
\end{theorem}

\Th~\ref{th:symmetry} states records in $o$ contributing equally to
leakage have the same Shapley value. Divergent Shapley values suggest divergent
leakage on the relevant program points. It ensures that all contributions are
awarded Shapley values.

% \vspace{-5pt}
\begin{theorem}[Dummy Player~\cite{shapley201617}]
    \label{th:dummy}
    If the $i$-th participant is a dummy player, i.e.,
    $\forall S \subseteq R^{o} \setminus \{i\}, \phi(o_{S \cup \{i\}}) - \phi(o_S)
    = \phi(o_{\{i\}}) - \phi(o_{\emptyset})$, then
    $\pi_{i}(\phi) = \phi(o_{\{i\}}) - \phi(o_{\emptyset})$.
\end{theorem}

\Th~\ref{th:dummy}, dubbed as ``Dummy Player,'' states that the information
leakage in one program point is not distributed to non-correlated points. In
sum, \Th~\ref{th:symmetry} and \Th~\ref{th:dummy} guarantee that the Shapley
value apportionment is \textbf{accurate}. Further, if $\phi(o_{\{i\}}) =
\phi(o_{\emptyset})$, we have the following theorem.

\vspace{-5pt}
\begin{subtheorem}[Null player~\cite{shapley201617}]
    \label{th:null}
    If the $i$-th participant has no contribution to any grand coalition game $\phi$,
    i.e., $\forall S \subseteq R^{o} \setminus \{i\}, \phi(o_{S \cup \{i\}}) = \phi(o_S)$,
    then $\pi_{i}(\phi) = 0$.
\end{subtheorem}

\Th~\ref{th:null} is one special case of \Th~\ref{th:dummy}, and it guarantees
that the Shapley value has \textbf{no false negative}. That is, program points
assigned with a zero Shapley value is guaranteed to not contribute to
information leakage.

\vspace{-5pt}
\begin{theorem}[Linearity~\cite{shapley201617}]
    \label{th:linearity}
    If two coalition games, namely $\phi$ and $\psi$, are combined, then
    $\pi_{i}(\phi + \psi) = \pi_{i}(\phi) + \pi_{i}(\psi)$ for
    $\forall i \in R^{o}$.
\end{theorem}

\Th~\ref{th:linearity} implies that if a secret has several independent
components, then the assigned Shapley value for each side channel record equals
to the linear sum of secrets leaked on this record from all components. 
For instance, let $\pi(\phi_{g})$ and $\pi(\phi_{a})$ be the leaked information
by recovering two privacy-related properties, ``gender'' and ``age,'' over
portrait photos. Since these two properties are independent, when considering
both, the newly-computed leakage $\pi(\phi)$ must equal the sum of
$\pi(\phi_{g})$ and $\pi(\phi_{a})$ according to \Th~\ref{th:linearity}. While
this property allows for fine-grained leakage analysis, we currently do not
separate a secret into independent components. We view this exploration as one
future work.

\vspace{-5pt}
\begin{theorem}[Uniqueness~\cite{shapley201617}]
    \label{th:uniqueness}
    The apportionment derived from Shapley value is the only one that
    simultaneously satisfies \Th~\ref{th:efficiency}, \Th~\ref{th:symmetry},
    \Th~\ref{th:dummy} \& \ref{th:null}, and \Th~\ref{th:linearity}.
\end{theorem}

\section{Responsible Disclosure}
\label{appx:confirmation}

\begin{table}[t]
  \vspace{-5pt}
  \caption{Statistics of reported vulnerabilities.}
  % \vspace{-5pt}
  \label{tab:report}
  \centering
\resizebox{0.9\linewidth}{!}{
  \begin{tabular}{c|c|c|c}
    \hline
       & Reported flaws & Answered & Acknowledged flaws \\
    \hline
    OpenSSL & 10 (functions)	            & 10             & 10             \\
    \hline
    MbedTLS & 62 (leakage sites)         & 62         & 62             \\
    \hline
    Libgcrypt & 5 (functions)         & 0         & 0             \\
    \hline
    Libjpeg & 2 (functions)         & 2         & 2            \\
    \hline
  \end{tabular}
  }
  % \vspace{-15pt}
\end{table}

We have reported our findings to developers. Since we find many vulnerabilities
in each cryptosystems and media software (see a complete list at~\cite{snapshot}),
we summarize representative cases and also clarify the leakage to developers to
seek prompt confirmation. Sometimes, as required, we further annotate the vulnerable
statements (e.g., for MbedTLS developers). \T~\ref{tab:report} lists the exact numbers
of vulnerable functions/program points we reported.

By the time of writing, OpenSSL developers have confirmed BIGNUM related
findings. MbedTLS developers also positively responded to our reports.
We are discussing with them the disclosure procedure. Libjpeg developers agreed
with our reported SDA/SCB cases but required PoC exploitations and patches to assess
cost vs. benefit. We did not receive response from Libgcrypt developers.

\end{appendix}

\end{document}